\documentclass[12pt]{article}
\usepackage{amsmath}
\usepackage{graphicx}
\usepackage{enumerate}
\usepackage{natbib}
\usepackage{url} 
\usepackage[active]{srcltx}
\usepackage{color}
\usepackage{colortbl} 
\usepackage{mathtools} 
\usepackage{array}
\usepackage{float}
\usepackage{booktabs}
\usepackage{mathrsfs}
\usepackage{multirow}
\usepackage{amsmath}
\usepackage{amsthm}
\usepackage{amssymb}
 \usepackage{bm}
\usepackage{graphicx}
\usepackage{setspace}
 \usepackage{enumerate}
\usepackage{environ}
\usepackage{soul}
\usepackage{tikz}
\usetikzlibrary{shapes, arrows.meta, positioning}
\usepackage[linesnumbered,ruled,vlined]{algorithm2e}
\usepackage{etoolbox} 

\AtBeginEnvironment{thebibliography}{\singlespacing} 
\patchcmd{\thebibliography}
  {\settowidth}
  {\setlength{\itemsep}{\baselineskip} 
   \setlength{\parskip}{0pt} 
   \setlength{\parsep}{0pt} 
   \settowidth}
  {}{}

\makeatletter
\bibpunct{(}{)}{;}{autheryear}{,}{;}
\allowdisplaybreaks
\providecommand{\tabularnewline}{\\}

\theoremstyle{plain}
\newtheorem{assumption}{\protect\assumptionname}
\theoremstyle{plain}

\theoremstyle{plain}

\theoremstyle{plain}

\theoremstyle{plain}

\@ifundefined{date}{}{\date{}}

\allowdisplaybreaks

\newtheorem{theorem}{Theorem}\newtheorem{corollary}{Corollary}\newtheorem{lemma}{Lemma}\newtheorem{remark}{Remark}

\providecommand{\lemmaname}{Lemma}
\providecommand{\remarkname}{Remark}
\providecommand{\theoremname}{Theorem}
\makeatother
\providecommand{\assumptionname}{Assumption}
\providecommand{\examplename}{Example}

\newcommand{\var}{\mathrm{var}}
\newcommand{\cov}{\mathrm{cov}}

\begin{document}
\def\spacingset#1{\renewcommand{\baselinestretch}%
{#1}\small\normalsize} \spacingset{1}
\newcommand{\blind}{1}
\if1\blind
{
  \title{\bf On the Role of Surrogates in Conformal Inference of Individual Causal Effects
 
}
  \author{Chenyin Gao$^{1}$, Peter B. Gilbert$^{2}$, Larry Han$^{2,3}$ \\ \\
  1 Department of Biostatistics, Harvard University \\
  2 Vaccine and Infectious Disease and Public Health Sciences Divisions, \\ Fred Hutchinson Cancer Center \\
  3 Department of Public Health and Health Sciences, \\ Northeastern University
  }
  \maketitle
} \fi

\if0\blind
{
  \bigskip
  \bigskip
  \bigskip
  \begin{center}
    {\LARGE\bf  On the Role of Surrogates in Conformal Inference of Individual Causal Effects}
\end{center}
  \medskip
} \fi

\bigskip

\begin{abstract}
    Learning the Individual Treatment Effect (ITE) is essential for personalized decision-making, yet causal inference has traditionally focused on aggregated treatment effects. While integrating conformal prediction with causal inference can provide valid uncertainty quantification for ITEs, the resulting prediction intervals are often excessively wide, limiting their practical utility. To address this limitation, we introduce \underline{S}urrogate-assisted \underline{C}onformal \underline{I}nference for \underline{E}fficient I\underline{N}dividual \underline{C}ausal \underline{E}ffects (SCIENCE), a framework designed to construct more efficient prediction intervals for ITEs. SCIENCE accommodates the covariate shifts between source data and target data and applies to various data configurations, including semi-supervised and surrogate-assisted semi-supervised learning. Leveraging semi-parametric efficiency theory, SCIENCE produces rate double-robust prediction intervals under mild rate convergence conditions, permitting the use of flexible non-parametric models to estimate nuisance functions. We quantify efficiency gains by comparing semi-parametric efficiency bounds with and without the surrogates. Simulation studies demonstrate that our surrogate-assisted intervals offer substantial efficiency improvements over existing methods while maintaining valid group-conditional coverage. Applied to the phase 3 Moderna COVE COVID-19 vaccine trial, SCIENCE illustrates how multiple surrogate markers can be leveraged to generate more efficient prediction intervals.
\end{abstract}

\noindent%
\vfill

\newpage

\section{Introduction}

Personalized medicine emphasizes the need for precise learning of individual treatment effects (ITEs) to guide patient-level decision-making. A central challenge in causal inference is the fundamental problem of missing data -- only one potential outcome is observed for each individual, while the others remain unobserved \citep{holland1986statistics}, rendering the ITE only partially identified. As a result, causal inference has traditionally focused on estimands like the average treatment effect (ATE) and the conditional average treatment effect (CATE), which provide insights at the population or subgroup level by averaging across individuals. However, these coarsened estimates may fail to capture individual-level heterogeneity that is critical for clinical decisions, especially for individuals who may respond differently to treatment.  This limitation is evident in contexts such as vaccine efficacy studies, where personalized vaccine recommendations are useful for guiding individual decisions. Two examples are the estimates for vulnerable immunocompromised individuals with a wide diversity of immunocompromised conditions, and estimates based on levels of antibodies against the targeted pathogen.

In such settings, surrogate outcomes -- biological markers, intermediate variables, or machine-learning-derived predictions that are easier to observe than the primary outcomes -- offer a promising avenue for improving the estimation of causal effects. Surrogate outcomes can often predict the primary outcome with reasonable accuracy and have been shown to enhance the efficiency of causal estimates, especially when the primary outcome is missing for a large proportion of individuals, e.g., the primary outcome may be invasive, costly, or take a long time to measure for many individuals. In vaccine studies, for instance, immune correlates of protection, such as neutralizing antibody titers, are now used as validated surrogates to predict vaccine efficacy for some contexts of use \citep{gilbert2022immune,gilbert2022covid}. 
 Despite these advantages, current statistical methods have yet to incorporate surrogate outcomes to enhance \textit{individual-level} causal inference. 
Our work addresses this gap by incorporating surrogate outcomes -- without imposing stringent conditions on their validity -- into a conformal inference framework. This approach enables more efficient uncertainty quantification of ITEs, advancing statistical theory and machine learning techniques in personalized medicine applications.

\subsection{Related literature on surrogates outcomes}

Since \citet{prentice1989surrogate} first proposed a definition and operational criterion for statistical surrogacy, several frameworks have been introduced for surrogate evaluation, including causal mediation \citep{robins1992identifiability}, principal stratification \citep{frangakis2002principal}, the proportion of treatment effect explained \citep{freedman1992statistical, parast2016robust, wang2020model, price2018estimation, han2022identifying, wang2023robust, gilbert2024surrogate}, and the meta-analytic framework \citep{buyse2000validation, burzykowski2005evaluation}. \citet{joffe2009related} and \citet{conlon2017links} characterized these methods broadly into the causal effects \citep{gilbert2008evaluating} and causal association \citep{li2010bayesian, alonso2016information} frameworks. For a comprehensive overview of surrogate evaluation methods in clinical trials, including recent developments, see  \citet{elliott2023surrogate}. This line of work aims to identify and validate surrogate outcomes as replacements for primary outcomes, enabling traditional or provisional approval of treatments through surrogate-based endpoints in clinical trials. Surrogate endpoint validation allows regulatory agencies, like the FDA, to accelerate approval processes for critical illnesses by relying on surrogate markers in place of target endpoints \citep{fda1992accelerated, fda2021}. However, surrogate markers are not always reliable; for example, drugs approved based on arrhythmia suppression (as a surrogate for mortality) later showed increased mortality in follow-up trials \citep{fleming1996surrogate}. To mitigate such risks, a range of surrogate validation criteria have been proposed, but these often impose stringent, untestable assumptions, such as the strong statistical surrogate assumption \citep{prentice1989surrogate} to fuse the target and source datasets, as in \cite{athey2019surrogate}.

In our approach, we avoid imposing such strict conditions, viewing surrogates as auxiliary information for the primary outcome rather than replacements. Our work builds upon a parallel line of work that focuses on utilizing surrogate outcomes to enhance the efficiency of treatment effect estimation on long-term or primary outcomes. \citet{athey2020combining} are motivated by combining experimental and observational data to estimate long-term treatment effects, where the experimental data includes only surrogates but the observational data includes surrogates and primary outcomes. \citet{chen2023semiparametric} studied semiparametric inference of the ATE in the settings considered by \citet{athey2020combining} and derived semiparametric efficiency bounds. \citet{cheng2021robust} and \citet{kallus2020role} studied efficient ATE estimation when combining a small number of primary outcome observations with many observations of the surrogates, without assuming strong statistical surrogacy. \citet{imbens2024long} explored the sequential structure of multiple surrogates and identified the average treatment effect on the primary outcomes in the presence of unmeasured confounders.

\subsection{Related literature on prediction-powered inference}

There is a growing literature on using black-box machine learning predictions as ``noisy surrogates" for the true outcome \citep{wang2020methods}.  One recent popular method is prediction-powered inference (PPI) \citep{angelopoulos2023}, which uses a small amount of gold standard labels and a large number of arbitrary machine learning predictions as surrogates. However, \citet{angelopoulos2023ppi++} showed that when the surrogates are inaccurate, the PPI interval can be worse than the standard interval using only the labeled data and proposed a fix called PPI++. \cite{miao2023assumption} showed that the PPI estimator is in general not statistically efficient and proposed an efficient influence function-based estimator. \cite{gan2023prediction} propose an alternative estimator that is guaranteed to be asymptotically no worse than the supervised counterpart and is equivalent to PPI++ \citep{angelopoulos2023ppi++} for a one-dimensional parameter. The general strategy in all of these works is to augment an estimator with an unbiased estimator of zero, leveraging its correlation with the estimation error to improve precision -- a well-established approach in missing data and semiparametric efficiency theory \citep{bickel1993efficient, robins1994estimation, robins1995analysis, hahn1998role}. Our work is connected to PPI in the sense that our proposed method can leverage machine-learned predictions as surrogates just as easily as it can use a validated surrogate, since we require no conditions on statistical surrogacy. The extent of efficiency gain will depend on the accuracy of the machine-learned prediction for the primary outcome.

However, all of these works, whether leveraging surrogate outcomes or machine-learned predictions as noisy surrogates, remain limited to population-level treatment effects, and it is not obvious how one could extend these methods to target the ITE.  As a result, while surrogates have been shown to improve ATE estimation, their potential to enhance individual-level causal inference remains unexplored.

\subsection{Related literature on conformal inference of counterfactuals}

Recent advances in conformal inference have shown promise in addressing the challenges of ITE estimation. \citet{lei2021conformal} provided the first connection between conformal prediction and causal inference, introducing a methodology to construct prediction sets for counterfactuals and ITEs. \citet{yang2024doubly} reformulated the prediction problem under covariate shift into a missing data problem and provided doubly robust and computationally efficient methods leveraging modern semiparametric efficiency theory. However, the prediction intervals produced for ITEs are typically too wide to be useful in practice. Neither of these works explores how incorporating surrogate outcomes could enhance the efficiency of the prediction intervals. 

\subsection{Contributions}

In this paper, we make several key contributions from both statistical and scientific perspectives. We consider different data configurations corresponding to the semi-supervised setting (Setting \ref{tab1}), the surrogate-assisted semi-supervised setting (Setting \ref{tab2}), and an intermediary setting where surrogates are observed in the source data only (Setting \ref{tab3}). In each setting, we derive the semi-parametric efficiency bounds for estimating the quantile of the non-conformity score, which is a necessary ingredient in constructing conformal prediction intervals (Theorems \ref{thm:EIF1} and \ref{thm:EIF-theta_C}) and establish the desired Probably Approximately Correct (PAC) asymptotic coverage (Theorem \ref{thm:DR_R}). Moreover, we analytically quantify the efficiency gains from observing surrogates and find that leveraging surrogates is beneficial when surrogates are reasonably predictive of the primary outcome (Corollaries \ref{coro:v1_reduction} and \ref{coro:v1_reduction_D0}).

From a methodological viewpoint, the proposed SCIENCE framework provides valid uncertainty quantification for causal estimands of interest, such as Individual Treatment Effects (ITEs), and more generally for any marginal contrast of the individual potential outcomes. The prediction intervals produced by SCIENCE are valid under minimal assumptions, offering both marginal coverage and group-conditional coverage guarantees. This ensures that the intervals are reliable not just on average but also across subpopulations. Furthermore, such asymptotic coverage is guaranteed for both the source data, where primary outcomes (i.e., one of the potential outcomes) are observed, and the target data, where primary outcomes are missing. By employing semiparametric efficiency theory and using cross-fitting \citep{chernozhukov2018double}, we allow for the use of any flexible machine learning estimators for estimating nuisance parameters, as long as they satisfy some generic convergence rate conditions. This approach enables us to achieve rate double robustness under relatively mild assumptions.

Our approach is validated through extensive simulations for both continuous and discrete outcomes, demonstrating significant efficiency gains in terms of shorter, more informative prediction intervals. An open-source R package is available for implementing our proposed methodology at \url{https://github.com/Gaochenyin/SurrConformalDR}. We analyze the phase 3 Moderna COVE COVID-19 vaccine efficacy trial, illustrating that incorporating surrogate markers, such as neutralizing and binding antibody levels at an early time point,  
 leads to more efficient prediction intervals. This analysis highlights the practical advantages of our methods in enhancing individual-level efficacy assessments.


\section{Preliminaries}
\subsection{Notation}
We consider a causal inference framework where individuals receive one of two treatments: $A \in \{0,1\}$, representing, for example, a vaccine ($A = 1$) or a placebo ($A = 0$). Let $X \in \mathcal{X} \subseteq \mathbb{R}^{d_X}$ denote the vector of baseline covariates, such as patient demographics and clinical characteristics. The primary outcome of interest is denoted $Y \in \mathbb{R}$, which could be quantitative or discrete, such as a clinical endpoint like the acquisition of symptomatic SARS-CoV-2 infection. Additionally, some individuals could have surrogate outcomes $S \in \mathcal{S} \subseteq \mathbb{R}^{d_S}$ measured, such as biomarkers (e.g., neutralizing antibody titers), that may be predictive of the primary outcome. The reason we treat the baseline covariates $X$ and the surrogate outcomes $S$ separately is that the surrogate outcomes can be post-treatment variables. Including these post-treatment variables $S$ as covariates in the analysis without additional assumptions may distort the causal relationship between $A$ and $Y$, as $S$ is observed only under the actual treatment but not under the counterfactual treatment.

Under the potential outcomes framework, we define $Y(a)$ and $S(a)$ as the potential primary outcome and surrogate outcome, respectively, that would be observed if an individual were assigned treatment $A=a$. We assume that each individual receives the treatment assigned in the study, so the observed outcomes are $(Y,S)=(Y(A), S(A))$. This assumption implies consistency and no interference between individuals; that is, one individual's treatment assignment does not affect another individual's outcomes. In certain contexts, such as vaccine studies, interference may occur. For example, the vaccination status of one individual could influence the infection risk of others through spillover effects \citep{hudgens2008toward, tchetgen2012causal}. Although important, this work will not address settings with interference.

We denote the units with observed primary outcomes as the source data and those without as the target data. Let $D \in \{0,1\}$ be an indicator of the data origin, with $D=1$ representing source data and $D=0$ representing target data. In summary, we observe a source dataset $\{(X_i,A_i,S_i,Y_i,D_i=1): i \in \mathcal{I}^s\}$ and a target dataset $\{(X_i,A_i,S_i,Y_i=\mathrm{NA},D_i=0): i \in \mathcal{I}^t\}$, where NA means ``not available" (missing), and $\mathcal{I}^s$ and $\mathcal{I}^t$ are the index sets for the source and target data, respectively. We denote $n_{D1} = |\mathcal{I}^s|$ and $n_{D0} = |\mathcal{I}^t|$ as the sample sizes for the two datasets, and $N = n_{D1} + n_{D0}$ as the total sample size. We assume that each data point $\mathcal{O}_i = (X_i, A_i, S_i, Y_i, D_i)$ is generated by coarsening an independent and identically distributed draw from a population $\mathcal{O}^* = (X,A,S(0),S(1),Y(0),Y(1),D)$ with an underlying joint distribution $\mathbb{P}$. The coarsening mechanism can be described by the map: 
$$
\mathcal{C}: (X,A, S(0),S(1),Y(0), Y(1),D)
\rightarrow 
(X,A,S(A), D \cdot Y(A) + (1-D) \cdot \textrm{NA}, D).
$$

\noindent Regular arithmetic operations apply to missing values, such that $D \cdot Y(A) + (1-D) \cdot \textrm{NA}$ is $Y(A)$ for source data $(D=1)$ and NA for target data $(D=0)$.

\subsection{Problem settings}

The semi-supervised setting (Setting \ref{tab1}) is considered as a baseline setting in which we observe $X,A,Y$ on units in the source data but only observe $X,A$ on units in the target data. The same data configuration was considered in \citet{zhang2022high} for semi-supervised estimation of heterogeneous treatment effects. In the context of transporting ATEs from a completed trial (source data) to a new population (target data), \citet{dahabreh2020extending} considered a slightly modified data configuration where the target data has only information on $X$ (rather than $X,A$). 

We then consider the surrogate-assisted semi-supervised setting (Setting \ref{tab2}), in which we observe $X,A,S,Y$ on units in the source data but only observe $X,A,S$ on units in the target data. Compared to Setting \ref{tab1}, we additionally observe $S$ on all units. The same data configuration was considered in \citet{kallus2020role} and \citet{cheng2021robust} in the context of estimation of ATEs. Setting \ref{tab2} represents a common application of a sufficiently-validated surrogate endpoint: bridging approval of a treatment $A=1$ with efficacy established from a completed phase 3 randomized trial with $X,A,S,Y$ measured (source data), to a new study population whose features may be shifted; for vaccines this widespread application is {\it immunobridging} \citep{krause2022making}.
The target dataset for the new population is a randomized trial with the surrogate $S$ used as the primary endpoint for the purpose of inferring a sufficient treatment effect on $Y$ for the target population. 
Immunobridging has been applied extensively for many vaccines, including for mRNA COVID-19 vaccines, for example for bridging efficacy against COVID-19 from adults to children \citep{walter2022evaluation}.

Finally, we consider an intermediate setting where $X,A,S,Y$ is observed in the source data, but only $X,A$ is available in the target data (Setting \ref{tab3}). We compare this setting to the baseline setting without surrogates (Setting~\ref{tab1}) in terms of the semiparametric efficiency lower bounds. Our analysis reveals that the efficiency lower bounds coincide under these two settings (Theorem~\ref{thm:EIF1}), meaning that when surrogates are observed only in the source data where primary outcomes are already available, they do not improve the efficiency for constructing the prediction intervals for the target data.

\begin{table}[!htbp]
\centering
\begin{minipage}{.3\linewidth}
\centering
\begin{tabular}{|ccccc|}
\toprule
Unit & $X$ & $A$ & $S$ & $Y$ \\ 
\midrule
\rowcolor{blue!20} 1 & \checkmark & \checkmark & ? & \checkmark \\ 
\rowcolor{blue!20} \vdots & \vdots & \vdots & \vdots & \vdots \\ 
\rowcolor{blue!20} $n_{D1}$ & \checkmark & \checkmark & ? & \checkmark \\ 
\rowcolor{red!20} $n_{D1}+1$ & \checkmark & \checkmark & ? & ? \\ 
\rowcolor{red!20} \vdots & \vdots & \vdots & \vdots & \vdots \\ 
\rowcolor{red!20} $n_{D1}+n_{D0}$ & \checkmark & \checkmark & ? & ? \\ 
\bottomrule
\end{tabular}
\caption{Semi-supervised setting\label{tab1}}
\end{minipage}
\hfill
\begin{minipage}{.3\linewidth}
\centering
\begin{tabular}{|ccccc|}
\toprule
Unit & $X$ & $A$ & $S$ & $Y$ \\ 
\midrule
\rowcolor{blue!20} 1 & \checkmark & \checkmark & \checkmark & \checkmark \\ 
\rowcolor{blue!20} \vdots & \vdots & \vdots & \vdots & \vdots \\ 
\rowcolor{blue!20} $n_{D1}$ & \checkmark & \checkmark & \checkmark & \checkmark \\ 
\rowcolor{red!20} $n_{D1}+1$ & \checkmark & \checkmark & \checkmark & ? \\ 
\rowcolor{red!20} \vdots & \vdots & \vdots & \vdots & \vdots \\ 
\rowcolor{red!20} $n_{D1}+n_{D0}$ & \checkmark & \checkmark & \checkmark & ? \\ 
\bottomrule
\end{tabular}
\caption{Surrogate-assisted semi-supervised setting\label{tab2}}
\end{minipage}
\hfill
\begin{minipage}{.3\linewidth}
\centering
\begin{tabular}{|ccccc|}
\toprule
Unit & $X$ & $A$ & $S$ & $Y$ \\ 
\midrule
\rowcolor{blue!20} 1 & \checkmark & \checkmark & \checkmark & \checkmark \\ 
\rowcolor{blue!20} \vdots & \vdots & \vdots & \vdots & \vdots \\ 
\rowcolor{blue!20} $n_{D1}$ & \checkmark & \checkmark & \checkmark & \checkmark \\ 
\rowcolor{red!20} $n_{D1}+1$ & \checkmark & \checkmark & ? & ? \\ 
\rowcolor{red!20} \vdots & \vdots & \vdots & \vdots & \vdots \\ 
\rowcolor{red!20} $n_{D1}+n_{D0}$ & \checkmark & \checkmark & ? & ? \\ 
\bottomrule
\end{tabular}
\caption{Intermediate setting\label{tab3}}
\end{minipage}
\end{table}

\section{Conformal inference under covariate shift}
\subsection{General framework}
Conformal inference is a powerful statistical framework for providing rigorous uncertainty quantification on individual predictions. Given independent and identically distributed source data $(W_i, Y_i)$ for $i=1,\cdots,N$, drawn from a source distribution $P_W\otimes P_{Y\mid W}$, where $W_i = (X_i, S_i)\in \mathcal{W}$ if surrogates $S_i$ are available, or $W_i = X_i$ otherwise, and a desired nominal coverage rate $1-\alpha \in (0,1)$, our goal is to construct prediction intervals $C(W)$ satisfying
$$
P_{(W_f,Y_f)\sim Q_W\otimes P_{Y\mid W}}(Y_f \in C(W_f))\geq 1-\alpha,
$$
where the pair $(W_f, Y_f)$ is drawn from the target distribution $Q_W \otimes P_{Y\mid W}$ and the probability is taken over the marginal distribution of the source data and the future observation $(X_f,Y_f)$. 

Here, $Q_W$ may differ from $P_W$, a scenario referred to as covariate shift when $W = X$ \citep{sugiyama2007covariate}. In our context where $W = (X,S)$, the joint distribution of covariates and surrogates can differ between the source and target datasets. If one is interested in a subset of the population where $W_f$ belongs to a subset $\mathcal{A}_W\subset \mathcal{W}$, the target distribution becomes $Q_W = P_{W\mid \mathcal{A}_W}$. In general, we can construct prediction intervals with valid group-conditional coverage by defining the target population as a specific subset characterized by $\mathcal{A}_W$. 

The prediction intervals $C(W)$ are determined by a one-dimensional function of $W$, denoted by $C(W;r_\alpha)$, and are constructed based on a non-conformity score $R(w,y)$. This score is a fixed non-stochastic function referred to as the non-conformity score \citep{vovk2005algorithmic}. For example, $R(w,y)$ can be the regression residual $R(w,y)=|y - \widehat{E}(y\mid w)|$ \citep{lei2018distribution} or $R(w,y) = \max\{
\widehat{q}_{\alpha/2}(w) - y, y-\widehat{q}_{1-\alpha/2}(w)
\}$, i.e., the conformalized quantile residual with estimated conditional quantiles $\widehat{q}_{\alpha/2}(\cdot)$ and $\widehat{q}_{1-\alpha/2}(\cdot)$ \citep{romano2019conformalized}. Specifically, for each value of $W$ and a threshold $r_\alpha$, the prediction interval is defined as $C(W;r_\alpha) = \{y \in \mathcal{Y} \mid R(W,y) \leq r_\alpha\}$. Given this definition, the event that the future response $Y_f$ falls within the prediction interval $C(W_f;r_\alpha)$ is equivalent to the event that the non-conformity score $R(W_f,Y_f)$ is below $ r_\alpha$, and their probabilities under the joint distribution $ Q_W\otimes P_{Y\mid W}$ are equal:
$$
P_{(W_f,Y_f)\sim Q_W\otimes P_{Y\mid W}}(Y_f \in C(W_f;r_\alpha)) = 
P_{(W_f,Y_f)\sim Q_W\otimes P_{Y\mid W}}(R(W_f, Y_f) \leq r_\alpha).
$$
If $r_{\alpha}$ is the smallest $(1-\alpha)$-quantile of $R(W_f, Y_f)$ for the target distribution $Q_W\otimes P_{Y\mid W}$, we have
$$
P_{(W_f,Y_f)\sim Q_W\otimes P_{Y\mid W}}(R(W_f, Y_f) \leq r_\alpha)\geq 1-\alpha,
$$
which holds irrespective of the choice of function $R(x,y)$. Therefore, our goal is to estimate the $(1-\alpha)$-quantile of $R(W_f, Y_f)$ from the source data following the distribution $P_W\otimes P_{Y\mid W}$ while $(W_f, Y_f) \sim Q_W\otimes P_{Y\mid W}$.

\subsection{Reformulation for individualized causal effects}
In this section, we first assume that the causal estimand $\theta_i$ does not depend on the joint distribution of potential outcomes and refer to it as the marginal contrast estimand. This class of estimands has been used in \cite{franks2020flexible} to bypass the need to specify the conditional copula for $Y(1)$ and $Y(0)$ when conducting sensitivity analysis. Under this condition, we can reformulate the covariate shift problem for the general conformal inference into two independent missing data problems, as seen in \cite{yang2024doubly}. 

\begin{assumption}\label{assum:estimand}
    Assume the individual causal estimand  $\theta_i=f(Y_i(1), Y_i(0))$ is a marginal contrast estimand, where $\theta_i$ can be completely characterized as a pre-specified transformation of the potential outcomes $Y_i(1)$ and $Y_i(0)$ separately, i.e., $\theta_i=u(Y_i(1)) - u(Y_i(0))$.
\end{assumption}
This class of transformations encompasses a wide range of estimands, including the ITE $Y(1)-Y(0)$ and the risk ratio under the logarithmic scale $\log(Y(1)) - \log(Y(0))$ for continuous outcomes. For ordinal outcomes, transformations $u(\cdot)$ can correspond to utility scores assigned to each level, often based on scientific domain knowledge, as observed in some stroke trials \citep{chaisinanunkul2015adopting, nogueira2018thrombectomy}. 

Next, the constructed prediction intervals $C_{\theta}(W_i;r_\alpha)$ for $\theta_i$ should satisfy:
\begin{align}
     P\{\theta_i\in C_{\theta}(W_i;r_\alpha)\}  &= 
     P(D_i=1)P(\theta_i \in C_\theta(W_i; r_\alpha)\mid D_i=1)\label{CF:D1} \\
     & + P(D_i=0)P(\theta_i \in C_\theta(W_i; r_\alpha)\mid D_i=0).\label{CF:D0}
\end{align}
For the term (\ref{CF:D1}), where one of the potential primary outcomes $Y(A)$ is observed for $D=1$, we only need to construct the prediction intervals for the counterfactuals $u\{Y(0)\}$ when $A=1$ and $u\{Y(1)\}$ when $A=0$:
\begin{align}
    &P(\theta_i \in C_\theta(W_i; r_\alpha)\mid D_i=1)\nonumber\\
    &= P(A_i = 1\mid D_i=1)P\{u\{Y_i(0)\}\in C_{0}(W_i;r_{\alpha,0})\mid A_i=1, D_i=1\}\nonumber\\
& + P(A_i = 0\mid D_i=1)P\{u\{Y_i(1)\}\in C_{1}(W_i;r_{\alpha,1})\mid A_i = 0, D_i=1\}\nonumber\\
& = P(A_i = 1\mid D_i=1)P(
R_{0,i} \leq r_{\alpha,0}\mid A_i=1, D_i=1)\label{CF:A0D1}\\
& + P(A_i = 0\mid D_i=1)P(
R_{1,i} \leq r_{\alpha,1}\mid A_i=0, D_i=1),\label{CF:A1D1} 
\end{align}
where $r_{\alpha,a}=\inf\{r:F_{R_a}(r\mid A=1-a, D=1)\geq 1-\alpha\}$ is the $(1-\alpha)$-quantile of the non-conformity score $R_{a,i} = R(W_i, u_a\{Y_i(a)\})$ conditional on $A=1-a$ and $D=1$. To connect it with the general conformal inference framework, the target distribution is $P(W,Y(a)\mid A=1-a, D=1)$, while the training data is distributed by $P(W, Y(a)\mid A =a, D=1)$ as $Y(a)$ is only observed when $A=a, D=1$; see Section \ref{sec:D1} for more details.

For term (\ref{CF:D0}), where both the potential primary outcomes are missing, our plan is to first create the data $(W_i, C_i)$ by
$$
C_i =
\begin{cases}
    u(Y_i) - C_0(W_i;r_{\alpha, 0}), & A_i=1, D_i=1\\
    C_1(W_i;r_{\alpha, 1}) - u(Y_i), & A_i=0, D_i=1,
\end{cases}
$$
where $C_i$ is the prediction intervals for $\theta_i$, which can be perceived as the pseudo-outcomes; Next, a secondary conformal inference is performed on the left- and the right-end point of $C_i$ to create the prediction intervals $C(W_i;r_{\gamma,C})$ satisfying:
$$
P(C_i \subset C(W_i;r_{\gamma,C})\mid D_i = 0) = P(R_{C,i}<r_{\gamma, C}\mid D_i=0) \geq 1-\gamma,
$$
where $R_{C,i}=R(W_i,C_i)$ is the non-conformity score for the prediction intervals. Thus, the asymptotic coverage is guaranteed as $P(\theta_i \not\in C(W_i; r_{\gamma, C})\mid D_i=0) \leq P(\theta_i \not\in C_i\mid D_i=0) + P(C_i \not\subset C(W_i; r_{\gamma, C})\mid D_i=0) \leq \alpha+\gamma$. Intuitively, this approach first imputes the missing primary outcomes for the target data with the pseudo-outcomes $C_i$, and then accounts for the uncertainty of the pseudo-outcomes by using $C(W; r_{\gamma, C})$. To connect it with the general conformal inference framework, the target distribution is $P(W, C\mid D=0)$ while the training data is distributed as $P(W,C \mid D=1)$; see Section \ref{sec:D0} for more details. 

\subsection{Assumptions}
We require the following assumptions to ensure the identification of $r_{\alpha, a}$ and $r_{\gamma, C}$.
\begin{assumption}\label{assump:overlap}
There exists some constant $0<c_0<1/2$ such that 
$c_0 \leq P(A=1\mid x) \leq 1-c_0$, $c_0 \leq  P(D=1\mid x, a)\leq 1-c_0$ for $a=0,1$ and any $x$ such that $f(x)>0$.
\end{assumption}
Assumption \ref{assump:overlap} states that each individual should have a probability of at least $c_0$ of being assigned to treatment or being included in the target data, conditional on any covariates with a positive likelihood of occurrence. This overlap assumption is common in the causal inference and missing data literature \citep{imbens2015causal}. In practice, this condition prevents situations where only certain types of individuals are treated or included in the target data, thereby enabling comparisons across treatments and datasets.
\begin{assumption}
\label{assump:unconfoundedness}
    $A \perp \{Y(a), S(a)\} \mid X$ for $a = 0, 1$.
\end{assumption}
Assumption \ref{assump:unconfoundedness} states that, conditional on the observed covariates $X$, the treatment assignment $A$ is independent of the potential primary outcomes $Y(a)$ and surrogates $S(a)$ for each treatment level $a$ in the combined data. This assumption is commonly referred to as unconfoundedness or ignorability of treatment assignment \citep{kallus2020role}. 

\begin{assumption}
\label{assump:external_validtiy}
    $D\perp S(a)\mid X, A$ for $a=0,1$.
\end{assumption}
Assumption \ref{assump:external_validtiy} states that the data origin indicator $D$ does not depend on the surrogates $S(a)$, given the observed variables $X$ and $A$. Therefore, the surrogates from combined data (if available) can be leveraged for efficiency gain.

\begin{assumption}
\label{assump:MAR}
$D\perp Y(a) \mid S(a), X, A$ for $a=0,1$.
\end{assumption}
Assumption \ref{assump:MAR} states that the data origin indicator $D$ is conditionally independent of the potential primary outcomes $Y(a)$ given the observed covariates $X$, the potential surrogate $S(a)$, and treatment assignment $A$. Since $D$ is equivalent to an indicator that $Y(a)$ is missing or not, Assumption \ref{assump:MAR} can be referred to as the missing at random (MAR) assumption in the missing data literature \citep{little2019statistical}, which implies that the observed variables fully explain any missingness in the primary outcomes. This is useful because it allows us to use the source data to infer information about the missing primary outcome in the target data, even in the presence of potential distributional shifts between data sources. Therefore, this condition is weaker than the missing completely at random (MCAR) condition assumed in the semi-supervised inference literature \citep{cheng2021robust, zhang2022high}, since it does not allow missingness to depend on any other variables. 

Below, we show that our assumptions are related to a similar set of identification assumptions in \cite{athey2020combining}.

\begin{lemma}
\label{lem:assumption}
    Under Assumptions \ref{assum:estimand} to \ref{assump:MAR}, we have (a)  $D\perp \{Y(a),S(a)\}\mid X$, (b) $A \perp \{Y(a), S(a)\} \mid X, D=0$, and (c) $A \perp \{Y(a), S(a)\} \mid X, D=1$ for $a = 0, 1$.
\end{lemma}

\begin{figure}[!htbp]
    \centering
    \includegraphics[width=0.85\linewidth]{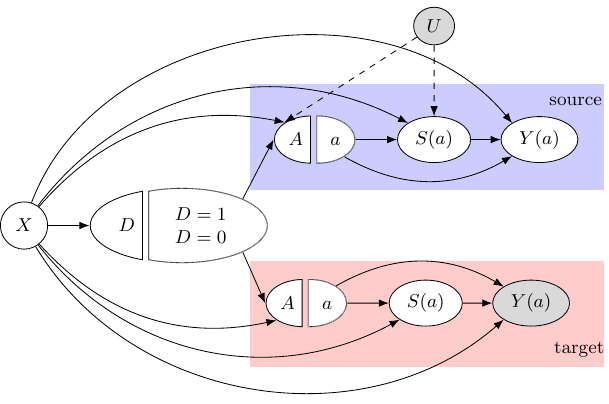}
    \caption{Single-world intervention graph for the source and target data with surrogates. Here, the shaded nodes, confounder $U$ and primary outcome $Y(a)$ of the target data, are unobserved. The dashed lines, that the unobserved confounder $U$ could affect treatment $A$ and surrogates $S(a)$ of the source data, are permitted in \cite{athey2020combining} but are not allowed for our problem setup.
    }
    \label{fig:SWIG}
\end{figure}

To align our problem setting with that in \cite{athey2020combining}, we could re-label our source data ($D = 1$) as their observational data and our target data ($D = 0$) as their experimental data. Then, Lemma \ref{lem:assumption} recovers the identification assumptions in \cite{athey2020combining}. In particular, Condition (a) corresponds to their external validity, and Condition (b) corresponds to their internal validity for the experimental data. Condition (c) is similar to their latent unconfoundedness for the observational data, i.e., $A\perp Y(a)\mid S(a), X, D=1$. Our assumptions additionally require unconfoundedness for the surrogates of the observational data, i.e. $A\perp S(a)\mid X, D=1$. This requirement is necessary to identify the non-conformity score $r_{\alpha, a}$ for $D=1$ in (\ref{CF:D1}). Since \cite{athey2020combining} primarily focuses on the identification across the combined population, this additional unconfoundedness condition is not necessary for their identification purposes; see Figure \ref{fig:SWIG} for a single world intervention graph for illustration.

\section{Conformal inference on the source data}\label{sec:D1}

\subsection{Identification}
\begin{theorem}(Identification formulas).
    Under Assumptions \ref{assum:estimand} to \ref{assump:MAR}, we have
        \begin{align*}
     &1-\alpha = P (R_a<{r}_{\alpha,a}\mid A = 1-a, D = 1)\\
     & = E_W\{P_Y (R_a<{r}_{\alpha,a}\mid A = a, D = 1, W)\mid A=1-a, D=1\}.
\end{align*} 
\end{theorem}
Thus, the identification formulas under Settings 1 -- 3 can be obtained accordingly:
    \begin{enumerate}
    \item[] 
    (Setting 1)
    $$
    1-\alpha = E_X\{P_Y (R_a<{r}_{\alpha,a}\mid A = a, D = 1, X)\mid A=1-a, D=1\},
    $$
    \item []
    (Setting 2)
    \begin{align*}
    &1-\alpha = E_W\{P_Y (R_a<{r}_{\alpha,a}\mid A = a, D = 1, W)\mid A=1-a, D=1\}\\
    & = E_X[E_S\{P_Y (R_a<{r}_{\alpha,a}\mid A = a, D = 1, S, X)\mid A = a, X\}\mid A=1-a, D=1],
    \end{align*}
    \item []
    (Setting 3)
    \begin{align*}
     &1-\alpha = E_W\{P_Y (R_a<{r}_{\alpha,a}\mid A = a, D = 1, W)\mid A=1-a, D=1\}\\
     & = E_X[E_S\{P_Y (R_a<{r}_{\alpha,a}\mid A = a, D = 1, S, X)\mid A = a, D=1, X\}\mid A=1-a, D=1] \\
     & = E_X\{P_Y (R_a<{r}_{\alpha,a}\mid A = a, D = 1, X)\mid A=1-a, D=1\}.
\end{align*}
The identification formula under Setting 3 reduces to the one under Setting 1, where the surrogates $S$ can only be leveraged for $D=1$, since they are missing for $D=0$.
    \end{enumerate}

\subsection{Semi-parametric efficiency analysis}
In this section, we derive the efficient influence functions (EIFs) for $r_{\alpha,1}$ under Settings 1 -- 3, and assess the potential efficiency gains from incorporating surrogates. The derivations for the EIFs of $r_{\alpha,0}$ follow analogously and are presented in Section \ref{subsec:thm_additional} of the Supplementary Materials. Key necessary model definitions are summarized in Table \ref{tab:notation_summary}.
\begin{table}[!htbp]
\centering
\caption{\label{tab:Key-assumptions} Lists of model definitions and details}
\vspace{0.5em}
\resizebox{\textwidth}{!}{
\begin{tabular}{p{.4\textwidth}p{.6\textwidth}}
\hline 
\textbf{Model} & \textbf{Definition}\tabularnewline
\hline 
$e_{A}(X)=P(A=1\mid X)$ & propensity score for treatment\\
$\pi_A(X)=\{1-e_A(X)\}/e_A(X)$ & inverse odds of treatment assignment \\
\hline
$e_D(X)=P(D=1\mid X)$ & propensity score for observing primary outcomes\\
$e_D(X,A)=P(D=1\mid X,A)$ & propensity score for observing primary outcomes within each group\\
$\pi_D(X)=\{1-e_D(X)\}/e_D(X)$ & inverse odds of observing primary outcomes \\
\hline
$\tilde{m}_a(r, X, S)$ \\
$=P(R_a<r\mid X,S,A=a,D=1)$ &  the conditional CDF of $R_a$ at $r$ with surrogates\\
$m_a(r, X)$ \\
$=P(R_a<r\mid X,A=a,D=1)$ & the conditional CDF of $R_a$ at $r$ without surrogates\\
\hline
$\tilde{m}_C(r, X, S)$ \\
$=P(R_C<r\mid X,S, D=1)$ & the conditional CDF of $R_C$ at $r$ with surrogates\\
$m_C(r, X)$\\
$=P(R_C<r\mid X,D=1)$ & the conditional CDF of $R_C$ at $r$ without surrogates\\
\hline
\end{tabular}}
\label{tab:notation_summary}
\end{table}

\begin{theorem}\label{thm:EIF1}
Under Assumptions \ref{assum:estimand} to \ref{assump:MAR}, the EIF $\psi_1^{(j)}$ for $r_{\alpha,1}$ under Setting $j$, for $j=S1, S2, S3$, is given, up to a proportionality constant, by
\begin{align*}
\psi_1^{(S1)}(r_{\alpha,1}, W;m,e_D,\pi_A)&=\psi_1^{(S3)}(r_{\alpha,1},W;m,e_D,\pi_A) =
    D(1-A)\{m_1(r_{\alpha,1}, X) - (1 - \alpha)\} \\
     &+ \frac{AD\pi_A(X)e_D(X,0)}{e_D(X,1)}
    \{\mathbf{1}(R_1<{r}_{\alpha,1}) - m_1(r_{\alpha,1}, X)\},
\end{align*}
and
\begin{align*}
\psi_1^{(S2)}(r_{\alpha,1}, W;m,\Tilde{m},e_D,\pi_A) &=  
    D(1-A)\{m_1(r_{\alpha,1}, X) - (1 - \alpha)\} \\
    & + A\pi_A(X) e_D(X,0)\{\Tilde{m}_1(r_{\alpha,1}, X,S) - m_1(r_{\alpha,1}, X)\}  \\
    & + \frac{AD\pi_A(X)e_D(X,0)}{e_D(X,1)}
    \{\mathbf{1}(R_1<{r}_{\alpha,1}) - \Tilde{m}_1(r_{\alpha,1}, X,S)\}.
\end{align*}
\end{theorem}
Theorem \ref{thm:EIF1} reveals that the EIFs for Settings 1 and 3 are the same. This means that when there is no surrogate or primary outcome information in the target data, having access to surrogates in the source data does not improve the efficiency of estimating $r_{\alpha,1}$ when we already have access to the primary outcomes in the source data. Intuitively, the surrogates provide no extra information if they are only observed for the units that already have primary outcomes observed. Essentially, surrogates cannot enhance estimation efficiency unless they help impute missing primary outcomes. In contrast, when surrogates are observed in both the source and target data, they can be leveraged to predict the missing primary outcomes in the target data, potentially improving estimation efficiency. 

Next, we formally quantify the efficiency gain in terms of the semi-parametric lower bounds from observing surrogates for both datasets as follows.
\begin{corollary}\label{coro:v1_reduction} 
Under Assumptions \ref{assum:estimand} to \ref{assump:MAR}, the efficiency gain from observing surrogates in both datasets (Setting 2) compared to only in the source data (Setting 1) in estimating $r_{\alpha,1}$ is given by
    \begin{align*}
        & V_1^{(S2)} - V_1^{(S1)} = 
        E \left[
        \frac{1 - e_D(X,1)}{e_D(X,1)}
        \frac{e_D^2(X,0)\{1-e_A(X)\}^2}{e_A(X)}
        \var \{\Tilde{m}_1(r_{\alpha, 1}, X,S)\mid X\}
        \right],
    \end{align*}
    where $ V_1^{(S1)} = 
    \var\{\psi_1^{(S1)}(r_{\alpha,1}, W;m,e_D,\pi_A)\}$ and $ V_1^{(S2)} = 
    \var\{\psi_1^{(S2)}(r_{\alpha,1}, W;m,\Tilde{m},e_D,\pi_A)\}$.
\end{corollary}
Corollary \ref{coro:v1_reduction} quantifies the efficiency gain in terms of the semiparametric efficiency lower bound for $r_{\alpha,1}$. Specifically, the efficiency gain depends on the predictiveness of surrogates for the primary outcomes. This is measured by the variance $\var \{\Tilde{m}_1(r_{\alpha, 1}, X,S)\mid X\}=\var \{P(R_1<r_{\alpha,1}\mid X, S, A=1, D=1)\mid X\}$, which reflects the additional variability in predicting $R_1$ that is captured by the surrogates $S$ beyond the covariates $X$. A higher variance means surrogates provide substantial information about the primary outcome, leading to greater efficiency gains when estimating $r_{\alpha,1}$.

Furthermore, the efficiency gain in estimating $r_{\alpha,1}$ is more pronounced when the proportion of primary outcomes observed is large for the control group but small for the treatment group. Intuitively, a higher proportion of primary outcomes observed in the treatment group indicates less missingness to address when estimating $r_{\alpha, 1}$, leading to less room for extra efficiency gain from leveraging surrogates. On the other hand, a higher proportion in the control group could amplify the efficiency gain by targeting a larger population as in (\ref{CF:A1D1}). As expected, the efficiency gain should be affected differently by $e_D(X,0)$ and $e_D(X,1)$ when estimating $r_{\alpha,0}$ (Corollary \ref{coro:v0_reduction}). Thus, reasonable gains in efficiency in estimating $\theta_i$ can be expected in settings with moderate levels of missingness for control and treatment groups.

\subsection{Implementation details}\label{sec:implementation}
\begin{figure}[!hbtp]
    \centering
    \includegraphics[width=0.75\linewidth]{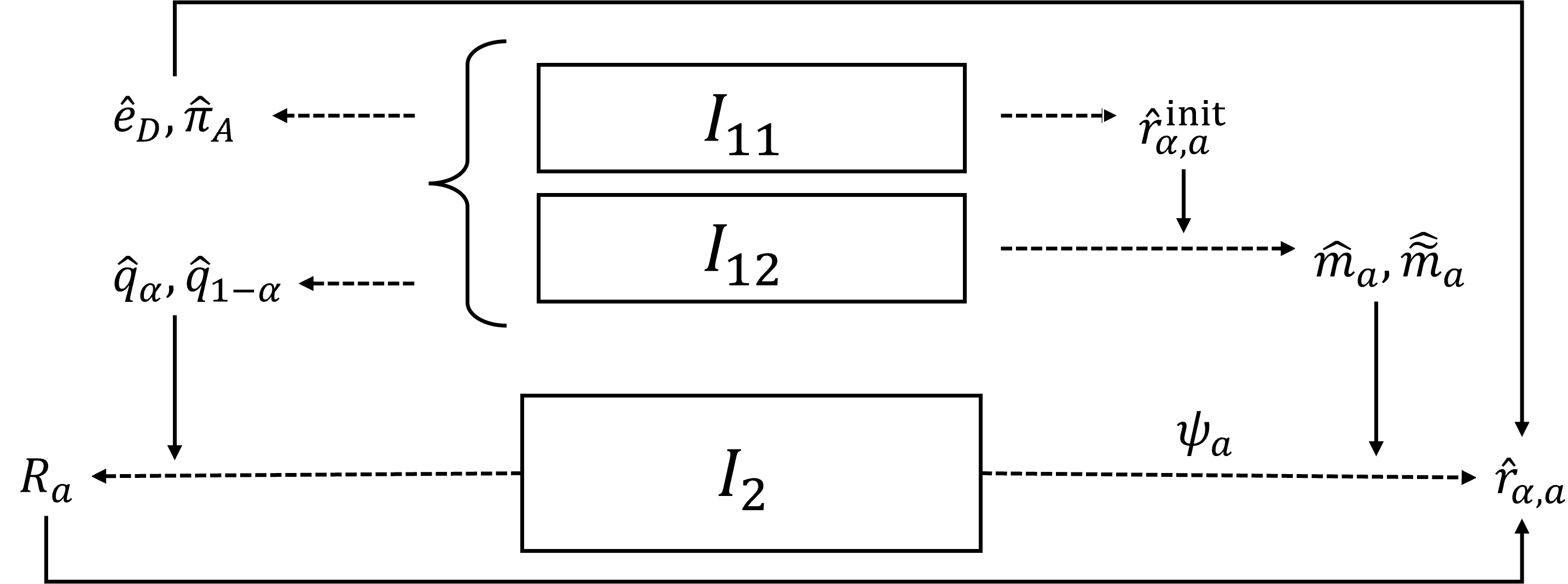}
    \caption{Schematic illustration for implementation of split conformal inference}
    \label{fig:flowchart}
\end{figure}

To construct the semi-parametric efficient estimators for $r_{\alpha, a}$, we adopt the split conformal inference strategy \citep{lei2014distribution}, which randomly splits the combined data into two folds, $\mathcal{I}_1$ and $\mathcal{I}_2$. The learning algorithms (e.g., the nuisance functions) are trained on the first fold, $\mathcal{I}_1$, while the prediction intervals are constructed using non-conformity scores on the second fold, $\mathcal{I}_2$. Details of the split conformal inference strategy are summarized in Figure \ref{fig:flowchart} for illustration.

For concreteness, we describe split conformal inference for $Y(a)$ using conformalized quantile residuals (CQR) on the source data. The implementation on the target data follows analogous logic. Let $\widehat{q}_{\alpha/2, a}(\cdot)$ and $\widehat{q}_{1-\alpha/2, a}(\cdot)$ be the quantile models for the primary outcomes $Y(a)$ trained on the first fold $\mathcal{I}_1$ with $A = a$. Additionally, the nuisance functions $\widehat{e}_A$ and $\widehat{e}_D$ are estimated on the first data fold, $\mathcal{I}_1$. To alleviate the computational burden of learning the entire conditional distribution $m_a(r, X)$ and $\tilde{m}_a(r, X)$ for infinitely many $r$, we adopt the localized debiased machine learning approach \citep{kallus2024localized}, where the first data fold, $\mathcal{I}_1$, is further divided into two sub-folds, $\mathcal{I}_{11}$ and $\mathcal{I}_{12}$. The first sub-fold, $\mathcal{I}_{11}$, is used to construct an initial estimator $\widehat{r}_{\alpha,a}^{\mathrm{init}}$ for $r_{\alpha,a}$. A natural initial estimator could be the weighted split-CQR estimator in \cite{lei2021conformal}. Then, the second sub-fold, $\mathcal{I}_{12}$, is used to estimate the nuisance functions $\widehat{m}_a(\widehat{r}_{\alpha,a}^{\mathrm{init}}, X)$ and $\widehat{\tilde{m}}_a(\widehat{r}_{\alpha,a}^{\mathrm{init}}, X,S)$ with any machine learning binary algorithm based on this single initial estimator $\widehat{r}_{\alpha,a}^{\mathrm{init}}$.

Next, the non-conformity scores $R_a$ are evaluated on the second fold, $\mathcal{I}_2$, with the estimated quantile models $\widehat{q}_{\alpha/2, a}(\cdot)$ and $\widehat{q}_{1-\alpha/2, a}(\cdot)$ as $R_{a,i} = \mathbf{1}(A_i=a)\max\{
\widehat{q}_{\alpha/2, a}(W_i) - Y_i, Y_i-\widehat{q}_{1-\alpha/2, a}(W_i)
\}$. Finally, we find the semi-parametric efficient estimators $\widehat{r}^{(S1)}_{\alpha, a}$ and $\widehat{r}^{(S2)}_{\alpha, a}$ on $\mathcal{I}_2$, defined as the smallest values satisfying 
$$
\sum_{i \in \mathcal{I}_2} \psi_a^{(S1)}(\widehat{r}_{\alpha,a}^{(S1)}, W_i;\widehat{m},\widehat{e}_D,\widehat{\pi}_A) \geq 0, \quad \sum_{i \in \mathcal{I}_2} \psi_a^{(S2)}(\widehat{r}_{\alpha,a}^{(S2)}, W;\widehat{m},\widehat{\Tilde{m}},\widehat{e}_D,\widehat{\pi}_A) \geq 0
$$
where $\widehat{m}$, $\widehat{\Tilde{m}}$, $\widehat{e}_D$, and $\widehat{\pi}_A$ are estimated from the first data fold $\mathcal{I}_1$.

\subsection{Asymptotic properties for coverage}
In this section, we establish asymptotic properties for the estimators of $r_{\alpha,1}$ and $r_{\alpha,0}$, which are important for constructing prediction intervals with the desired coverage level $1-\alpha$. First, we establish the connection between the asymptotic coverage probability $P(R_a<r_{\alpha,a}\mid A=1-a,D=1)$ and the EIF for $r_{\alpha, a}$.
\begin{lemma}\label{lem:Dr}
    Under Assumptions \ref{assum:estimand} to \ref{assump:MAR}, the following holds for any EIF $\psi_a(r_{\alpha, a}, W)$ under Settings 1 -- 3:
    $$
    P(R_a<r_{\alpha,a}\mid A=1-a,D=1) =  1-\alpha + \frac{E\{\psi_a(r_{\alpha, a}, W)\}}{P(A=1-a, D=1)}.
    $$
\end{lemma}
Lemma \ref{lem:Dr} states that the asymptotic coverage probability deviates from the nominal level $1-\alpha$ by a term proportional to the expected value of the EIF. We now list some regularity conditions for the nuisance functions that will be useful in establishing the asymptotic coverage probability for the estimators of $r_{\alpha, a}$:
\begin{enumerate}
    \item [(A1)] The estimated functions $\widehat{\pi}_A(X)$, $\widehat{e}_D(X, A)$, $\widehat{m}_a(r, X)$, and $\widehat{\tilde{m}}_a(r, X,S)$ are bounded, i.e., there exist $\pi_0$, $\underline{e}_0$, $\overline{e}_0$, $m_0$, and $\tilde{m}_{0}$, such that $|\widehat{\pi}_A(X)| \leq \pi_{0}$, $\underline{e}_{0}\leq |\widehat{e}_D(X, A)|\leq \overline{e}_{0}$, $|\widehat{m}_a(r, X)|\leq m_{0}$, and $|\widehat{\tilde{m}}_a(r, X,S)|\leq \tilde{m}_{0}$.

    \item [(A2)] The estimators $\widehat{m}_a(r, X)$ and $\widehat{\tilde{m}}_a(r, X,S)$ are non-decreasing function of $r$.
\end{enumerate}
Following the split conformal inference strategy in Section \ref{sec:implementation}, the desired asymptotic coverage is guaranteed for the source data in Theorem \ref{thm:DR_R}.
\begin{theorem}\label{thm:DR_R}
    Under Assumptions \ref{assum:estimand} to \ref{assump:MAR} and regularity conditions (A1) and (A2), there exist some constants $C_0$ and $C_1$ for any $\delta>0$, such that 
\begin{align*}
    &P(
R_a \leq \widehat{r}_{\alpha,a}^{(S1)}\mid A_i=1-a, D_i=1)\geq (1-\alpha)\\
& - C_0
\frac{\pi_{0}\overline{e}_{0}\underline{e}_{0}^{-1}(1+ m_{0})}{P(A=1-a, D=1)}\sqrt{\frac{\log(1/\delta) + 1}{|\mathcal{I}_2|}}\\
   &-C_1\left\{ \frac{\|\widehat{\pi}_A(X) - {\pi}_A(X)\|}{P(A=1-a, D=1)}\cdot \sup_r\|\widehat{m}_a(r,X) - m_a(r,X)\|\right. \\
   &\left. +\frac{\|\widehat{e}_D(X,A) - e_D(X,A)\|}{P(A=1-a, D=1)}\cdot \sup_r\|\widehat{m}_a(r,X) - m_a(r,X)\|\right\},
\end{align*}
and
\begin{align*}
   &P(
R_a \leq \widehat{r}_{\alpha,a}^{(S2)}\mid A_i=1-a, D_i=1) \geq (1-\alpha) \\
& - C_0\frac{\pi_{0}\overline{e}_{0}(\underline{e}_{0}^{-1}+ m_{0} +\underline{e}_{0}^{-1}\tilde{m}_{0})}{P(A=1-a, D=1)}\sqrt{\frac{\log(1/\delta) + 1}{|\mathcal{I}_2|}}\\
   &-C_1\left\{\frac{\|\widehat{\pi}_A(X) - {\pi}_A(X)\|}{P(A=1-a, D=1)}\cdot \sup_r\|\widehat{m}_a(r,X) - m_a(r,X)\|\right.\\
   & \left.+ \frac{\|\widehat{e}_D(X,A) - e_D(X,A)\|}{P(A=1-a, D=1)}\cdot \sup_r\|\widehat{m}_a(r,X) - m_a(r,X)\|\right.\\
& \left.+ \frac{\|\widehat{e}_D(X,A) - e_D(X,A)\|}{P(A=1-a, D=1)}\cdot \sup_r\|\widehat{\Tilde{m}}_a(r, X,S) - 
{\Tilde{m}}_a(r,X,S)\|\right\}
\end{align*}
hold with probability greater than $1-\delta$.
\end{theorem}


Theorem \ref{thm:DR_R} provides a coverage guarantee of $1-\alpha$ for the prediction intervals of $u\{Y(1)\}$ and $u\{Y(0)\}$. Note that the slack for the coverage guarantee is the sum of two parts: the first part is induced by approximating the expectation with its empirical version, which scales as $O(N^{-1/2})$, by splitting the data into two folds $\mathcal{I}_1$ and $\mathcal{I}_2$ of similar size; the second part is induced by the product bias from the estimation of the nuisance functions, which is negligible if either \(\|\widehat{\pi}_A(X) - {\pi}_A(X)\|+\|\widehat{e}_D(X,A) - e_D(X,A)\|=o_p(1)\) or \(\sup_r\|\widehat{m}_a(r,X) - m_a(r,X)\|+\sup_r\|\widehat{\Tilde{m}}_a(r,X,S) - {\Tilde{m}}_a(r,X,S)\|=o_p(1)\). Such property is known as rate double robustness, which means that the asymptotic coverage is robust to small perturbations in the nuisance functions, such that their estimation errors affect the coverage error only in second-order terms \citep{chernozhukov2018double}. 

Hence, the constructed prediction intervals $C_\theta(W_i; \widehat{r}_\alpha)$ by the SCIENCE framework achieve the desired coverage level asymptotically in the source data under Settings 1 to 3, up to a negligible term vanishing in probability:
\begin{align*}
     &P(\theta_i \in C_\theta(W_i; \widehat{r}_\alpha)\mid D_i=1)\\
     &=P(A_i = 1\mid D_i=1)P(
R_0 \leq \widehat{r}_{\alpha,0}\mid A_i=1, D_i=1)\\
& + P(A_i = 0\mid D_i=1)P(
R_1 \leq \widehat{r}_{\alpha,1}\mid A_i=0, D_i=1) \geq (1-\alpha) + o_p(1),
\end{align*}
which is often referred to as Probably Approximately Correct (PAC) coverage in the statistical literature \citep{krishnamoorthy2009statistical}.

\section{Conformal inference on the target data}\label{sec:D0}
In this section, we address the challenging task of constructing valid prediction intervals for the target data where primary outcomes are unobserved ($D=0$). Building upon the nested conformal inference framework proposed by \cite{lei2021conformal}, we develop a two-step procedure to achieve this goal. 

\begin{figure}[!htbp]
    \centering
\includegraphics[width=0.85\linewidth]{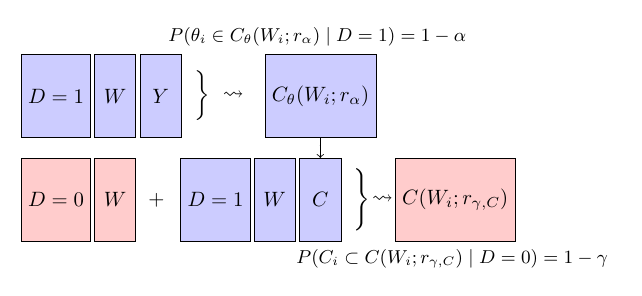}
    \caption{Schematic illustration for constructing the nested prediction intervals $C(W;r_{\alpha,C})$ for target data $D=0$}
    \label{fig:sketch_RC}
\end{figure}

First, we generate a dataset $(W_i, C_i)$ by constructing prediction intervals $C_i=C_\theta(W_i;r_\alpha)$ for each individual in the source data ($D=1$). These intervals $C_i$ are designed to cover the causal estimand $\theta_i$ with probability at least $1-\alpha$ conditional on the source data, using the methods detailed in Section \ref{sec:D1}. Next, we treat the constructed intervals $C_i$ as pseudo-outcomes and perform a secondary conformal inference to find a prediction interval $C(W;r_{\gamma,C})$ that nests $C_i$ with high probability; $C(W;r_{\gamma,C})$ is called the nested prediction interval. Specifically, we aim to determine $r_{\gamma,C}$ such that
\begin{equation}\label{eq
} P\big( C_i \subset C(W_i; r_{\gamma,C}) \mid D_i = 0 \big) = P\big( R_{C,i} \leq r_{\gamma,C} \mid D_i = 0 \big) = 1 - \gamma, \end{equation} 
where $R_{C,i}=R(W_i,C_i)$ is the non-conformity score measuring the discrepancy between the pseudo-outcome $C$ and its prediction based on $W$. Figure \ref{fig:sketch_RC} provides a schematic illustration of this nested procedure, and an example of constructing $R_C$ is given in Remark \ref{rmk:exampleC}. 

\begin{remark}
\label{rmk:exampleC}
    Consider the causal estimand as the ITE, $\theta_i = Y_i(1) -Y_i(0)$. For the source data $D=1$, we construct prediction intervals $C_i=C_{\theta}(W_i;r_{\alpha})$ as follows
$$
C_i =
\begin{cases}
    &Y_i - C_0(W_i;r_{\alpha,0}),\quad A_i = 1, D_i=1\\
    &C_1(W_i;r_{\alpha,1}) - Y_i,\quad A_i= 0, D_i=1.
\end{cases}
$$
Next, we compute the non-conformity scores $R_{C,i}$ using the conformalized quantile residuals $R_{C,i} =  
  \max\{\widehat{m}^L(W_i)-C_i^L, C_i^R - \widehat{m}^L(W_i)\}$, where $(C_i^L, C_i^R)$ are the lower and upper endpoints of $C_i$, and $(\widehat{m}^L(W_i), \widehat{m}^R(W_i))$ are estimates of the conditional means of $(C_i^L, C_i^R)$ given $W_i$.
\end{remark}

Having obtained the non-conformity scores $R_C$ for individuals in the source data, we aim to extend this information to the target data $D=0$. Under Assumptions \ref{assump:external_validtiy} and \ref{assump:MAR}, we can relate the distribution of non-conformity scores in the target data to that in the source data through the following identification formula for the quantile $r_{\gamma,C}$:
\begin{align*}
    1-\gamma = P(R_C < r_{\gamma,C} \mid D = 0) = E_W\left\{P(R_C < r_{\gamma,C} \mid W, D = 1)\mid D=0\right\},
\end{align*}
where the desired coverage probability $1-\gamma$ in the target data is ensured, adjusted for differences in the covariate distributions between the two datasets.

The identification formula serves as the foundation for estimating $r_{\gamma,C}$ and ensures that the constructed prediction intervals $C(W;r_{\gamma,C})$ achieve nominal coverage in the target data. Moreover, it motivates derivation of the EIF for $r_{\alpha,C}$, which we present in Theorem \ref{thm:EIF-theta_C}. 
\begin{theorem}\label{thm:EIF-theta_C}
    Under Assumptions \ref{assump:overlap}-\ref{assump:unconfoundedness}, let $r_{\gamma,C}$ be the $(1-\gamma)$-th quantile of the non-conformity scores $R_C$. The EIF for $r_{\gamma,C}$ is, up to a proportionality constant,
    for Setting 2,
    \begin{align*}
        &\psi_C^{(S2)}(r_{\gamma,C}, W; m_C, \Tilde{m}_C,e_D, \pi_A) = (1-D)\{m_C(r_{\gamma,C},X)- (1-\gamma)\} \\
    &+\{1-e_D(X)\}\{\Tilde{m}_C(r_{\gamma,C},X,S) - m_C(r_{\gamma,C},X)\} + 
    D\pi_D(X)\{\mathbf{1}(R_C<r_{\gamma,C}) - \Tilde{m}_C(r_{\gamma,C},X,S)\},
    \end{align*}
    and for Settings 1 and 3,
    \begin{align*}
    &\psi_C^{(S1)}(r_{\gamma,C},W; m_C,e_D, \pi_A) =\psi_C^{(S3)}(r_{\gamma, C}, W; m_C,e_D, \pi_A) \\
    &= (1-D)\{m_C(r_{\gamma,C},X)- (1-\gamma)\} + 
    D\pi_D(X)\{\mathbf{1}(R_C<r_{\gamma,C}) - m_C(r_{\gamma,C},X)\}.
\end{align*}
\end{theorem}
Analogously, we could quantify the optimal efficiency gain from leveraging surrogates for estimating $r_{\gamma,C}$, which we formalize in Corollary \ref{coro:v1_reduction_D0}.
\begin{corollary}\label{coro:v1_reduction_D0} 
Under Assumptions \ref{assump:overlap}-\ref{assump:unconfoundedness}, the efficiency gain from using surrogates in estimating $r_{\gamma,C}$ in the target data is
    \begin{align*}
        V_C^{(S1)} - V_C^{(S2)} = 
        E \left[
        \pi_D(X)\{1-e_D(X)\}^2
        \var \{\Tilde{m}_C(r_{\gamma,C},X,S)\mid X\}
        \right],
    \end{align*}
    where  $V_C^{(S1)} = 
    \var\{\psi_C^{(S1)}(r_{\gamma,C}, W;m_C,e_D,\pi_A)\}$, $V_C^{(S2)} = 
    \var\{\psi_C^{(S2)}(r_{\gamma,C}, W;m_C,\Tilde{m}_C,e_D,\pi_A)\}$.
\end{corollary}
Corollary \ref{coro:v1_reduction_D0} shows that the efficiency gain from incorporating surrogates depends positively on the variance of $\tilde{m}_C(r_{\gamma,C},X,S)$ given $X$. This variance captures the additional predictiveness of the surrogates for the non-conformity scores $R_C$ given $X$. 

To estimate $r_{\gamma,C}$ in practice, define $\widehat{r}_{\gamma,C}^{(S1)}$ and $\widehat{r}_{\gamma,C}^{(S2)}$ to be the smallest values satisfying
$$
\sum_{i \in \mathcal{I}_2} \psi_C^{(S1)}(\widehat{r}_{\gamma,C}^{(S1)}, W_i;\widehat{m}_C,\widehat{e}_D,\widehat{\pi}_A) \geq 0, \quad \sum_{i \in \mathcal{I}_2} \psi_C^{(S2)}(\widehat{r}_{\gamma,C}^{(S2)}, W;\widehat{m}_C,\widehat{\Tilde{m}}_C,\widehat{e}_D,\widehat{\pi}_A) \geq 0,
$$
where the nuisance functions are estimated on the first data fold $\mathcal{I}_1$. Similar to Theorem \ref{thm:DR_R}, we can show that under both Setting 1 (or 3) and Setting 2, the coverage of the constructed prediction intervals satisfies
\begin{align*}
   P(\theta_i &\notin C(W_i; \widehat{r}_{\gamma,C})\mid D_i=0)\\
    &\leq 
    P(\theta_i \notin C_i \mid D_i=0) + 
    P(C_i \not\subset C(W_i; \widehat{r}_{\gamma,C}) \mid D_i=0) \leq \alpha + \gamma + o_p(1),
\end{align*}
demonstrating that the probability of the true causal parameter $\theta_i$ not being contained in the final prediction interval $C(W_i; \widehat{r}_{\gamma,C})$ is bounded above by $\alpha+\gamma$ up to a negligible term. The nested prediction intervals are designed to 
adapt jointly to the uncertainty of the lower and upper endpoints of $C_i$, ensuring that the overall coverage is at least $1-(\alpha+\gamma)$. Thus, we can show that under both Setting 1 (or 3) and Setting 2, $$P(\theta_i \in C_\theta(W_i; \widehat{r}_{\gamma,C})\mid D_i=0)
 \geq 1-(\alpha + \gamma),$$ 
 holds asymptotically, which guarantees the asymptotic coverage.

\section{Simulation studies}

In this section, we demonstrate the performance of SCIENCE and showcase its efficiency gain under various settings via simulation studies. We begin by considering a data-generation procedure similar to that in \citet{kallus2020role}.

Specifically, we first generate baseline covariate \(X \in \mathbb{R}^2\) following a multivariate normal distribution \(X \sim N(0, I_2)\) for the entire population, with total sample sizes \(N = 3000, 5000, 10000\). Next, we randomly select a diminishing proportion \(P(D=1) = N^{-1/4}\) of the entire population as the source data (\(D=1\)), while the remaining data constitute the target data (\(D=0\)). The treatment variable is generated according to a logistic regression model: \(P(A=1 \mid X) = 1/\{1+\exp(-\alpha_A - \sum_{j}\eta_j X_j)\}\), where \(\eta_1=\eta_2 = -1/2\) and \(\alpha_A\) is adaptively chosen to ensure an average treatment rate of 0.5. We then simulate potential surrogate outcomes \(S(a) \in \mathbb{R}^2\) by \(S(a) \sim N((-1)^{a+1}, \sigma_S^2 I_2)\), where \(\sigma_S = 1, 5, 10, 50\). Finally, we generate potential primary outcomes \(Y(a)\) by:
$$
Y(a) = (-1)^{a+1} + 
\frac{(-1)^a}{2}
\frac{\sum_{j=1}^2 S_j(a)}{5}+
\sum_{j=1}^2 \beta_j X_j + \epsilon, \quad 
\epsilon \sim N(0, 1),
$$
where \(\beta_1 = \beta_2 = 1\). One can observe that the increase in the explained variance of \(Y(a)\) by the surrogates \(S(a)\), adjusting for the baseline covariates, is \(\sigma_S^2/100\). Thus, the variance \(\sigma_S^2\) quantifies how predictive the surrogates are for the primary potential outcomes \(Y(a)\), after accounting for the baseline covariates. Moreover, the surrogates can be considered as important as the covariates when $\sigma_S = 10$, which we treat as the benchmark scenario for the subsequent model evaluation.

We are interested in evaluating the properties of the prediction intervals for the ITEs over the combined data, as well as separately for the target and source data. In our analysis, \(X\), \(A\), \(S(A)\), and \(Y(A)\) are observed for the source data, while only \(X\), \(A\), and \(S(A)\) are observed for the target data. Therefore, we can calculate the proposed semi-parametric efficient estimators both with and without access to the surrogates in the target data. The non-conformity scores \(R_a\) and \(R_C\) are constructed using conformalized quantile residuals from the quantile regression model available in the \texttt{quantreg} package, which fits the 2.5\% and 97.5\% conditional quantiles of \(Y(a)\). The propensity scores and other nuisance functions required for finding the efficient estimators are fitted using \texttt{SuperLearner}, with Random Forest and generalized linear models as base learners. We split 75\% of the data as the first fold \(\mathcal{I}_1\) for model training, as suggested by \cite{sesia2020comparison}, and use the remaining data to construct the prediction intervals. Furthermore, we compare the proposed methods against the weighted CQR (WCQR) proposed in \cite{lei2021conformal}. 

\begin{figure}[!ht]
    \centering
    \includegraphics[width=.9\linewidth]{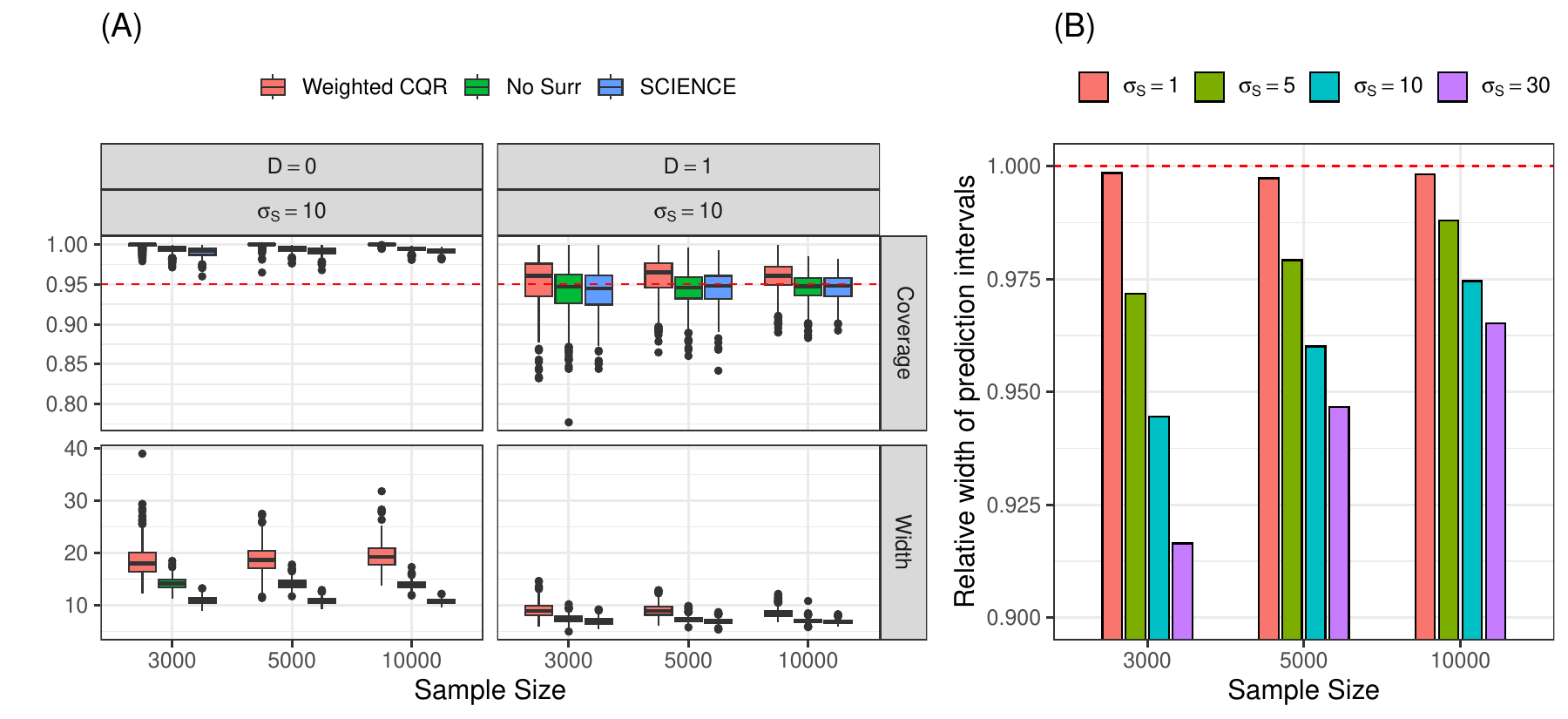}
    \caption{(A) Empirical coverage and width of the 95\% prediction intervals,  when $\sigma_S = 10$ and (B) average relative width of the $95\%$ prediction intervals (`SCIENCE' versus `No Surr') when $\sigma_S = 1,5, 10, 30$ for ITE across 500 replicates.}
    \label{fig:rst_combined}
\end{figure}

Figure \ref{fig:rst_combined}(A) presents the empirical coverage and average width of the prediction intervals for the ITE across 500 Monte Carlo simulations. While all methods are guaranteed to cover the ITE, we note that nested conformal inference for the target data tends to be overly conservative when primary outcomes are missing, a phenomenon also observed in \cite{lei2021conformal}. As expected, both semi-parametric efficient estimators produce shorter intervals than the WCQR, while still maintaining valid coverage. Among all methods, SCIENCE achieves the shortest intervals. The efficiency gains from using surrogates are further illustrated in Figure \ref{fig:rst_combined}(B) for $\sigma_S=1,5,10,30$. The benefits are measured by the average relative width of the prediction intervals compared to those obtained without using surrogates. As demonstrated by Corollaries \ref{coro:v1_reduction} and \ref{coro:v1_reduction_D0}, the improvement from using surrogates is more pronounced as $\sigma_S$ increases. This is because, with higher \(\sigma_S\), surrogates account for more variation of the primary outcomes,  leading to better predictiveness when constructing prediction intervals for the ITE.

\subsection{Group-conditional performance for continuous outcomes}
\begin{figure}[!h]
    \centering
    \includegraphics[width=.9\linewidth]{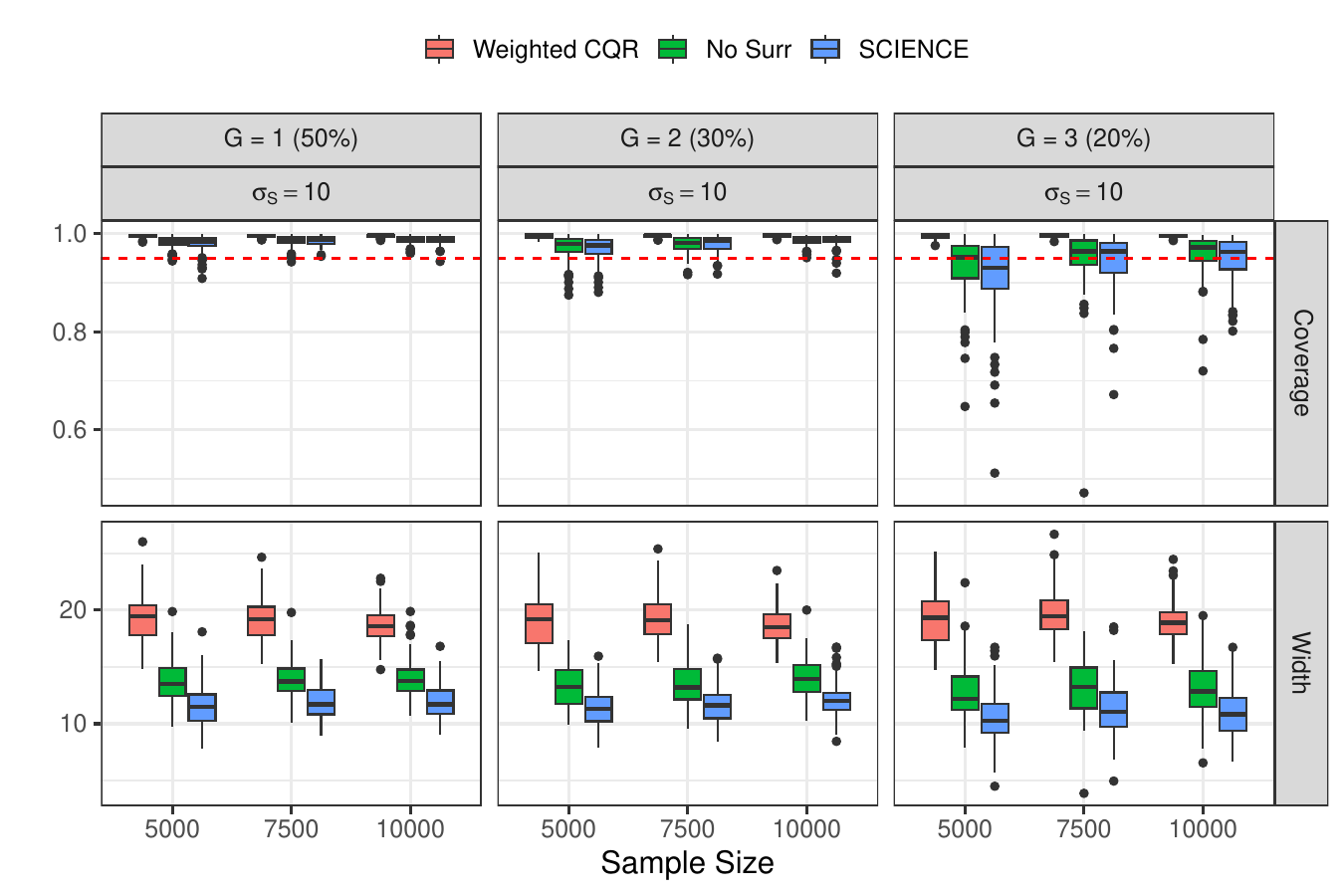}
    \caption{Empirical coverage and width of the 95\% prediction intervals for the ITE of the combined data when $\sigma_S = 10$, conditional on the group variable $G$, across 500 replicates.}
    \label{fig:rst1wR}
\end{figure}

In this section, we evaluate the group-conditional performance of the SCIENCE framework for continuous outcomes; more simulation studies of group-conditional performance for categorical outcomes are provided in Section \ref{more:sims} of the Supplementary Material. We follow data-generation procedures similar to those in the previous section, except that we additionally generate the group variable $G$ with three levels from a multinomial distribution. Specifically, $P(G=1 \mid X) \propto \exp\left(-\alpha_{G,1}-\sum_j X_j/2\right)$, $P(G=2 \mid X) \propto \exp\left(-\alpha_{G,2}-X_1-X_2/2\right)$, and $P(G=3 \mid X) \propto \exp\left(-\alpha_{G,3}-X_1/2-X_2\right)$, where $\alpha_R = (\alpha_{G,1}, \alpha_{G,2},\alpha_{G,3})$ is adaptively chosen to ensure the proportion of each group is around 0.5, 0.3, and 0.2, respectively. The generation of the treatment variable $A$ and the potential primary outcomes $Y(a)$ are modified to include the effect of group variable: $P(A=1 \mid X, G) = 1/\{1+\exp(-\alpha_A - \sum_{j}\eta_j X_j + G)\}$, and
$$
Y(a) = (-1)^{a+1} + 
\frac{(-1)^a}{2}
\frac{\sum_{j=1}^2 S_j(a)}{5}+
\sum_{j=1}^2 \beta_j X_j + G + \epsilon, \quad 
\epsilon \sim N(0, 1).
$$
Figure \ref{fig:rst1wR} presents the empirical coverage and average width of the prediction intervals for the ITE of the combined data when $\sigma_S = 10$, conditional on the group variables $G$. All methods achieve valid $95\%$ conditional coverage within each group and SCIENCE produces the shortest intervals -- on average, 48\% and 22\% shorter than the WCQR and efficient estimator without surrogates, respectively.

\section{Real data analysis}

We analyzed the Moderna COVE phase 3 COVID-19 vaccine efficacy trial that randomized adults to receive two doses of mRNA-1273 vaccination or placebo at Day 1 and Day 29, conducting a per-protocol analysis that included participants who received both injections without specified protocol violations.  Participants were randomly sampled into the immunogenicity subcohort for measurement of antibody markers at the Day 1, Day 29, and Day 57 visits, using stratified Bernoulli random sampling. 

Our analysis focused on a cohort of $257$ individuals who were baseline seropositive for SARS-CoV-2, defined as being RNA or serology positive. This selection was motivated by the fact that individuals without prior SARS-CoV-2 infection would be seronegative at baseline, leading to the structural condition that $Y_i(0)=0$ for all individuals in the placebo arm, i.e., with no SARS-CoV-2 antigen exposure, the antibody levels must be zero. 
For completeness, we present results for baseline seronegative individuals in Supplement \ref{more:real-data}.

Among participants, 127 were placebo recipients ($A=0$), and 130 received the two-dose mRNA-1273 vaccine $(A=1)$.  We designated the target population as underrepresented minorities $n_{D0} = 110$ $(42.8\%)$, defined as Blacks or African Americans, Hispanics or Latinos, American Indians or Alaska Natives, Native Hawaiians, and other Pacific Islanders. The source population of non-minorities $n_{D1} = 147$ $(57.2\%)$ included all other races with observed race (Asian, Multiracial, White, Other) and observed ethnicity not being Hispanic or Latino.

We controlled for the same confounders considered in previous analyses, including an indicator of being at high risk for COVID-19, an age-comorbidity stratification variable, and a standardized risk score built using machine learning on placebo arm participants. The standardized risk score was built based on a prespecified list of baseline covariates, including age, sex, BMI, enrollment time, and other variables \citep{gilbert2022immune}.

We considered two surrogate markers: (i) Day 29 pseudo-neutralizing antibody readout, the $\textrm{log}_{10}$ serum inhibitory dilution 50\% titer (nAb-ID50), and (ii) Day 29 binding IgG spike antibody $\textrm{log}_{10}$ concentration, which is cheaper and easier to measure. The primary outcome $Y$ was the Day 57 $\textrm{log}_{10}$ nAb-ID50 titer, selected because it has been widely used as a primary endpoint in immunobridging studies.  The parameter of interest is the ITE $Y(1)-Y(0)$.

We used a 75-25 split for training and testing, as in the simulation studies. To account for the variability from data splitting, we repeated the above procedure 100 times and summarized the average coverage and widths of prediction intervals of the observed outcomes for different methods. Our results show that SCIENCE produces prediction intervals with nominal coverage across the combined target and source populations for various miscoverage levels $\alpha \in \{0.05, 0.1, 0.2, 0.3, 0.4\}$ (see Table~\ref{maintab:coverage_and_width_overall}). In terms of average prediction interval width for the ITE, summarized in the same table, WCQR yields the widest intervals. SCIENCE generates shorter average intervals than the efficient method without surrogates. Using the Day 29 nAb-ID50 titer as a surrogate results in narrower intervals compared to the Day 29 binding antibody titer alone, with the shortest intervals achieved when both Day 29 markers are used. As expected, prediction intervals for the target population are slightly wider due to the nested conformal inference procedure, and intervals increase in width as $\alpha$ decreases. Detailed results for the source and target populations are provided in Table~\ref{maintab:cp_width_source_target} of the Supplementary Material.

\begin{table}[htbp]
\label{maintab:coverage_and_width_overall}
\centering
\caption{Baseline SARS-CoV-2 seropositive participants: Average coverage and prediction interval width across the combined population of target and source test points over $100$ repetitions at different miscoverage levels $\alpha$, comparing WCQR \citep{lei2021conformal}, efficient method without surrogates (No Surr), SCIENCE using Day 29 binding spike antibody level as the surrogate (Binding), SCIENCE using Day 29 nAb-ID50 titer as the surrogate (nAb-50), and SCIENCE using both surrogates (Both).}
\vspace{0.5em}
\resizebox{\textwidth}{!}{
\begin{tabular}{c|ccccc|ccccc}
\toprule
\textbf{$\alpha$} & \multicolumn{5}{c|}{\textbf{Coverage $\times 100\%$}} & \multicolumn{5}{c}{\textbf{Prediction Interval Width}} \\
\cmidrule(lr){2-6} \cmidrule(lr){7-11}
 & \textbf{WCQR} & \textbf{No Surr} & \textbf{SCIENCE} & \textbf{SCIENCE} & \textbf{SCIENCE} & \textbf{WCQR} & \textbf{No Surr} & \textbf{SCIENCE} & \textbf{SCIENCE} & \textbf{SCIENCE} \\
  &  & & \textbf{(Binding)} & \textbf{(nAb-50)} & \textbf{(Both)} & &  & \textbf{(Binding)} & \textbf{(nAb-50)} & \textbf{(Both)} \\
\midrule
0.05 & 95.5 & 95.0 & 95.2 & 94.8 & 95.3 & 2.07 & 1.96 & 1.88 & 1.85 & 1.72 \\
0.10 & 91.3 & 89.0 & 90.1 & 89.9 & 90.1 & 1.69 & 1.55 & 1.48 & 1.47 & 1.44 \\
0.20 & 81.5 & 80.3 & 80.9 & 79.9 & 80.9 & 1.28 & 1.23 & 1.19 & 1.18 & 1.15 \\
0.30 & 66.1 & 70.0 & 72.0 & 70.3 & 71.7 & 1.05 & 1.02
& 1.02 & 1.00 & 0.98 \\
0.40 & 59.4 & 59.2 & 61.3 & 60.5 & 60.9 & 0.89 & 0.88 & 0.86 & 0.85 & 0.83 \\
\bottomrule
\end{tabular}
 }
\label{tab:cp_width}
\end{table}

\begin{figure}[!htbp]
    \centering
    \includegraphics[width=.8\linewidth]{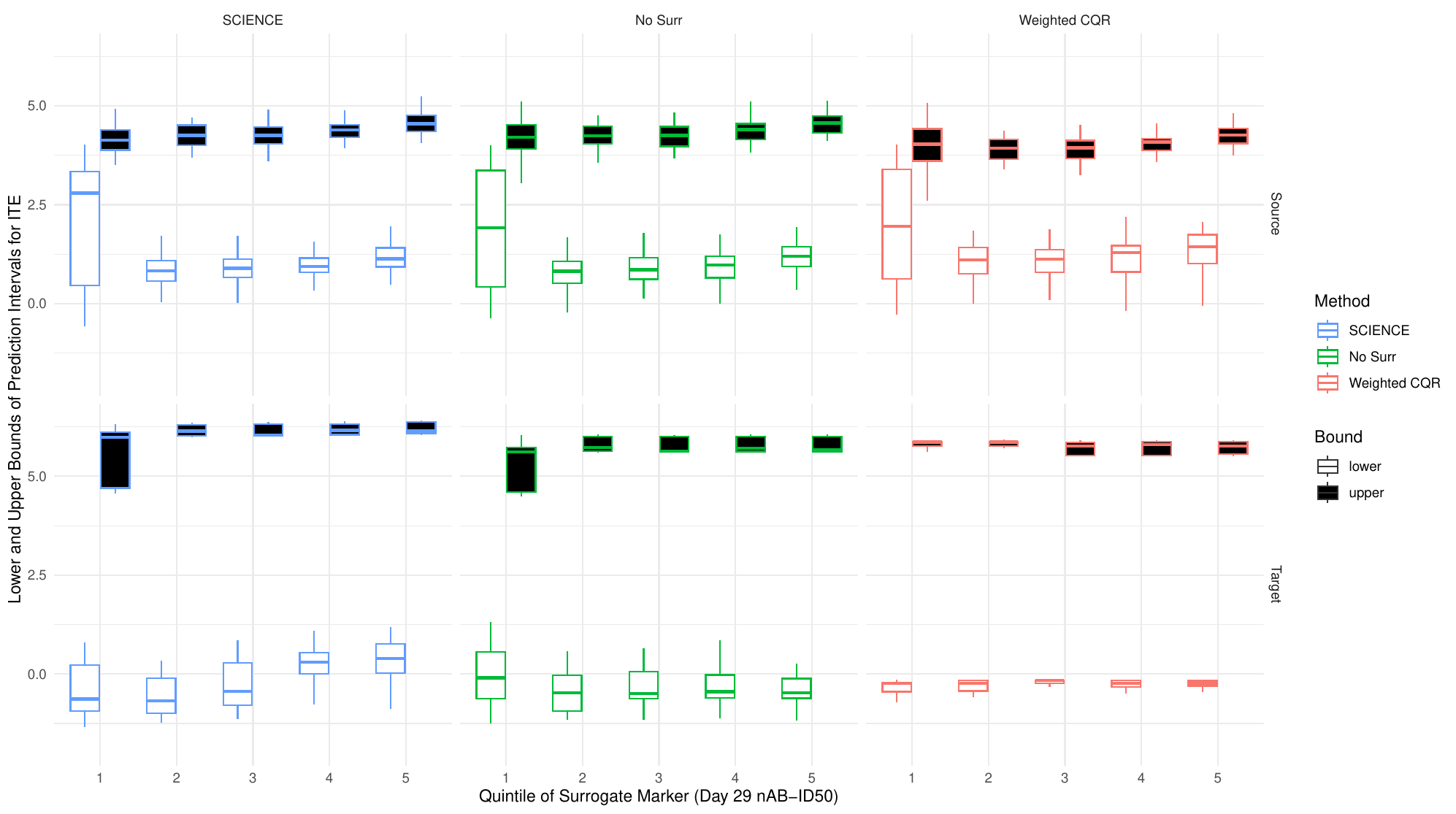}
    \caption{Lower and upper bounds of the $90\%$ prediction intervals for test points on the source and target populations for the ITE on the Day 57 nAb-ID50 primary outcome  ($Y_i(1) - Y_i(0)$) across quintiles of Day 29 nAb-ID50 surrogate marker titer; Quintile 1: $(0.0828,0.501]$, Q2: $(0.501,0.944]$, Q3: $(0.944, 1.291]$, Q4: $(1.291, 1.627]$, Q5: $(1.627, 4.068]$ in units IU50/ml}
    \label{fig:nAbsurr}
\end{figure}

To visualize the efficiency gain from incorporating surrogate markers, we plotted the lower and upper bounds of the 90\% prediction intervals for the ITE on log$_{10}$ Day 57 neutralizing antibody titer.
The prediction intervals were evaluated across five levels of surrogate marker values: very low, low, medium, high, and very high, with quintiles computed by pooling data from both vaccine and placebo arms. Within each quintile, we plotted the bounds of the prediction intervals for test points in both the target and source populations. We focused on Setting 2, incorporating the nAb-ID50 surrogate, while also considering Setting 1, represented by the Weighted CQR and No Surr estimators. As shown in Figure~\ref{fig:nAbsurr}, in the source population, the lower bounds of the prediction intervals were generally positive across all values of the surrogate, indicating a favorable vaccine effect relative to placebo. Furthermore, the prediction intervals shifted to higher values for individuals with higher surrogate marker levels, suggesting that higher Day 29 nAb-ID50 titer predicted a larger vaccine effect on the Day 57 titer. Notably, SCIENCE yielded the narrowest prediction intervals, demonstrating the efficiency gain achieved by incorporating surrogate information. In the target population, the lower bound of the ITE was consistently below zero for Weighted CQR. However, SCIENCE identified a positive ITE for a large proportion of individuals.

\section{Discussion}
In this paper, we proposed a novel surrogate-assisted conformal inference framework (SCIENCE) for constructing more efficient individualized prediction intervals. Our method is broadly applicable, accommodating scenarios where primary outcomes may or may not be observed. By leveraging semiparametric efficient estimation, SCIENCE is capable of handling distributional shifts between source and target populations, achieving the desired PAC (probably approximately correct) asymptotic coverage property, where coverage bias is a product of estimation errors. Consequently, SCIENCE ensures nominal coverage rates, provided that flexible non-parametric models used for nuisance function estimation meet the required rate of convergence conditions.

While our approach achieves group-conditional coverage over pre-specified subgroups, future research could extend these guarantees to more comprehensive and formally developed notions of fairness. Moreover, prediction intervals are not the only form of uncertainty quantification. An intriguing avenue for future work would be the exploration of conformal predictive distributions, which output a probability distribution over the response space rather than a fixed prediction interval, allowing for potentially more informative uncertainty quantification \citep{vovk2017nonparametric}. Another area of interest is class-conditional coverage, where coverage is guaranteed for specific outcome classes \citep{ding2024class}. This extension could enhance the applicability of our method in settings where outcome-dependent coverage is crucial for decision-making, thereby improving the interpretability and reliability of individualized prediction intervals across different outcome groups. Finally, post-selection inference remains an essential consideration for practical applications. Iteratively choosing and refining the miscoverage rate $\alpha$ can be valuable when an initially aggressive choice (i.e., small $\alpha$) results in prediction intervals that are too wide to be useful. Investigating adaptive strategies for re-selecting $\alpha$ to balance interval width and coverage validity could improve the adoption of conformal inference methods for precision medicine and beyond.

\section{Acknowledgments}
The authors thank the COVE study participants and COVE study team, other Coronavirus Prevention Network colleagues, and Moderna colleagues. The authors also thank Matteo Bonvini for helpful discussions and feedback that enhanced the manuscript.
This research was supported in part by the Administration for Strategic Preparedness and Response, Biomedical Advanced Research and Development Authority Contracts No. 75A50120C00034 (P3001 study) and No. 75A50122C00013 (Immune Assays) and the National Institute of Allergy and Infectious Diseases (NIAID) of the National Institutes of Health through grant UM1AI068635. The findings and conclusions herein are those of the authors and do not necessarily represent the views of the Department of Health and Human Services or its components. The content is solely the responsibility of the authors and does not necessarily represent the official views of the National Institutes of Health. The content of this publication does not necessarily reflect the views or policies of the Department of Health and Human Services, nor does mention of trade names, commercial products, or organizations imply endorsement by the U.S. Government.
\bibliography{main}

@article{dahabreh2020extending,
  title={Extending inferences from a randomized trial to a new target population},
  author={Dahabreh, Issa J and Robertson, Sarah E and Steingrimsson, Jon A and Stuart, Elizabeth A and Hernan, Miguel A},
  journal={Statistics in Medicine},
  volume={39},
  pages={1999--2014},
  year={2020},
  publisher={Wiley Online Library}
}

@article{angelopoulos2023,
author = {Anastasios N. Angelopoulos  and Stephen Bates  and Clara Fannjiang  and Michael I. Jordan  and Tijana Zrnic },
title = {Prediction-powered inference},
journal = {Science},
volume = {382},
pages = {669-674},
year = {2023},
doi = {10.1126/science.adi6000},
URL = {https://www.science.org/doi/abs/10.1126/science.adi6000},
eprint = {https://www.science.org/doi/pdf/10.1126/science.adi6000}}

@article{angelopoulos2023ppi++,
  title={Ppi++: Efficient prediction-powered inference},
  author={Angelopoulos, Anastasios N and Duchi, John C and Zrnic, Tijana},
  journal={arXiv Preprint arXiv:2311.01453},
  year={2023}
}

@article{miao2023assumption,
  title={Assumption-lean and data-adaptive post-prediction inference},
  author={Miao, Jiacheng and Miao, Xinran and Wu, Yixuan and Zhao, Jiwei and Lu, Qiongshi},
  journal={arXiv Preprint arXiv:2311.14220},
  year={2023}
}

@article{gan2023prediction,
  title={Prediction De-Correlated Inference},
  author={Gan, Feng and Liang, Wanfeng},
  journal={arXiv Preprint arXiv:2312.06478},
  year={2023}
}

@book{bickel1993efficient,
  title={Efficient and adaptive estimation for semiparametric models},
  author={Bickel, Peter J and Klaassen, Chris AJ and Bickel, Peter J and Ritov, Ya’acov and Klaassen, J and Wellner, Jon A and Ritov, YA'Acov},
  volume={4},
  year={1993},
  publisher={Springer}
}

@article{wang2020methods,
  title={Methods for correcting inference based on outcomes predicted by machine learning},
  author={Wang, Siruo and McCormick, Tyler H and Leek, Jeffrey T},
  journal={Proceedings of the National Academy of Sciences},
  volume={117},
  pages={30266--30275},
  year={2020},
  publisher={National Acad Sciences}
}

@article{ding2024class,
  title={Class-conditional conformal prediction with many classes},
  author={Ding, Tiffany and Angelopoulos, Anastasios and Bates, Stephen and Jordan, Michael and Tibshirani, Ryan J},
  journal={Advances in Neural Information Processing Systems},
  volume={36},
  year={2024}
}

@article{hahn1998role,
  title={On the role of the propensity score in efficient semiparametric estimation of average treatment effects},
  author={Hahn, Jinyong},
  journal={Econometrica},
  pages={315--331},
  year={1998},
  publisher={JSTOR}
}

@article{robins1994estimation,
  title={Estimation of regression coefficients when some regressors are not always observed},
  author={Robins, James M and Rotnitzky, Andrea and Zhao, Lue Ping},
  journal={Journal of the American Statistical Association},
  volume={89},
  pages={846--866},
  year={1994},
  publisher={Taylor \& Francis}
}

@article{robins1995analysis,
  title={Analysis of semiparametric regression models for repeated outcomes in the presence of missing data},
  author={Robins, James M and Rotnitzky, Andrea and Zhao, Lue Ping},
  journal={Journal of the American Statistical Association},
  volume={90},
  pages={106--121},
  year={1995},
  publisher={Taylor \& Francis}
}

@article{zhang2022high,
  title={High-dimensional semi-supervised learning: in search of optimal inference of the mean},
  author={Zhang, Yuqian and Bradic, Jelena},
  journal={Biometrika},
  volume={109},
  pages={387--403},
  year={2022},
  publisher={Oxford University Press}
}

@article{joffe2009related,
  title={Related causal frameworks for surrogate outcomes},
  author={Joffe, Marshall M and Greene, Tom},
  journal={Biometrics},
  volume={65},
  pages={530--538},
  year={2009},
  publisher={Oxford University Press}
}

@article{conlon2017links,
  title={Links between causal effects and causal association for surrogacy evaluation in a Gaussian setting},
  author={Conlon, Anna and Taylor, Jeremy and Li, Yun and Diaz-Ordaz, Karla and Elliott, Michael},
  journal={Statistics in Medicine},
  volume={36},
  pages={4243--4265},
  year={2017},
  publisher={Wiley Online Library}
}

@article{li2010bayesian,
  title={A Bayesian approach to surrogacy assessment using principal stratification in clinical trials},
  author={Li, Yun and Taylor, Jeremy MG and Elliott, Michael R},
  journal={Biometrics},
  volume={66},
  pages={523--531},
  year={2010},
  publisher={Oxford University Press}
}

@article{fda1992accelerated,
  title={Accelerated approval of new drugs for serious or life-threatening illnesses},
  author={FDA},
  journal={Federal Register},
  volume={57},
  pages={58942–58960},
  year={1992}
}

@misc{fda2021,
  author = {FDA},
  title = {Accelerated Approval Program},
  year = 2021,
  howpublished = {\url{https://www.fda.gov/drugs/nda-and-bla-approvals/accelerated-approval-program}},
  note = {Accessed: 2024-10-31}
}

@article{walter2022evaluation,
  title={Evaluation of the BNT162b2 Covid-19 vaccine in children 5 to 11 years of age},
  author={Walter, Emmanuel B and Talaat, Kawsar R and Sabharwal, Charu and Gurtman, Alejandra and Lockhart, Stephen and Paulsen, Grant C and Barnett, Elizabeth D and Mu{\~n}oz, Flor M and Maldonado, Yvonne and Pahud, Barbara A and others},
  journal={New England Journal of Medicine},
  volume={386},
  pages={35--46},
  year={2022},
  publisher={Mass Medical Soc}
}

@book{vovk2005algorithmic,
  title={Algorithmic learning in a random world},
  author={Vovk, Vladimir and Gammerman, Alexander and Shafer, Glenn},
  volume={29},
  year={2005},
  publisher={Springer}
}

@inproceedings{vovk2017nonparametric,
  title={Nonparametric predictive distributions based on conformal prediction},
  author={Vovk, Vladimir and Shen, Jieli and Manokhin, Valery and Xie, Min-ge},
  booktitle={Conformal and probabilistic prediction and applications},
  pages={82--102},
  year={2017},
  organization={PMLR}
}

@article{sugiyama2007covariate,
  title={Covariate shift adaptation by importance weighted cross validation.},
  author={Sugiyama, Masashi and Krauledat, Matthias and M{\"u}ller, Klaus-Robert},
  journal={Journal of Machine Learning Research},
  volume={8},
  year={2007}
}

@article{hudgens2008toward,
  title={Toward causal inference with interference},
  author={Hudgens, Michael G and Halloran, M Elizabeth},
  journal={Journal of the American Statistical Association},
  volume={103},
  pages={832--842},
  year={2008},
  publisher={Taylor \& Francis}
}

@article{tchetgen2012causal,
  title={On causal inference in the presence of interference},
  author={Tchetgen, Eric J Tchetgen and VanderWeele, Tyler J},
  journal={Statistical Methods in Medical Research},
  volume={21},
  pages={55--75},
  year={2012},
  publisher={SAGE Publications Sage UK: London, England}
}

@article{gilbert2022immune,
  title={Immune correlates analysis of the mRNA-1273 COVID-19 vaccine efficacy clinical trial},
  author={Gilbert, Peter B and Montefiori, David C and McDermott, Adrian B and Fong, Youyi and Benkeser, David and Deng, Weiping and Zhou, Honghong and Houchens, Christopher R and Martins, Karen and Jayashankar, Lakshmi and others},
  journal={Science},
  volume={375},
  pages={43--50},
  year={2022},
  publisher={American Association for the Advancement of Science}
}

@article{holland1986statistics,
  title={Statistics and causal inference},
  author={Holland, Paul W},
  journal={Journal of the American Statistical Association},
  volume={81},
  pages={945--960},
  year={1986},
  publisher={Taylor \& Francis}
}

@techreport{athey2019surrogate,
  title={The surrogate index: Combining short-term proxies to estimate long-term treatment effects more rapidly and precisely},
  author={Athey, Susan and Chetty, Raj and Imbens, Guido W and Kang, Hyunseung},
  year={2019},
  institution={National Bureau of Economic Research}
}

@article{chen2023semiparametric,
  title={Semiparametric estimation of long-term treatment effects},
  author={Chen, Jiafeng and Ritzwoller, David M},
  journal={Journal of Econometrics},
  volume={237},
  pages={105545},
  year={2023},
  publisher={Elsevier}
}

@article{lei2021conformal,
  title={Conformal inference of counterfactuals and individual treatment effects},
  author={Lei, Lihua and Cand{\`e}s, Emmanuel J},
  journal={Journal of the Royal Statistical Society Series B: Statistical Methodology},
  volume={83},
  pages={911--938},
  year={2021},
  publisher={Oxford University Press}
}

@article{yang2024doubly,
  title={Doubly robust calibration of prediction sets under covariate shift},
  author={Yang, Yachong and Kuchibhotla, Arun Kumar and Tchetgen Tchetgen, Eric},
  journal={Journal of the Royal Statistical Society Series B: Statistical Methodology},
  pages={qkae009},
  year={2024},
  publisher={Oxford University Press US}
}

@article{lei2018distribution,
  title={Distribution-free predictive inference for regression},
  author={Lei, Jing and G’Sell, Max and Rinaldo, Alessandro and Tibshirani, Ryan J and Wasserman, Larry},
  journal={Journal of the American Statistical Association},
  volume={113},
  pages={1094--1111},
  year={2018},
  publisher={Taylor \& Francis}
}

@article{romano2019conformalized,
  title={Conformalized quantile regression},
  author={Romano, Yaniv and Patterson, Evan and Candes, Emmanuel},
  journal={Advances in Neural Information Processing Systems},
  volume={32},
  year={2019}
}

@article{kallus2020role,
  title={On the role of surrogates in the efficient estimation of treatment effects with limited outcome data},
  author={Kallus, Nathan and Mao, Xiaojie},
  journal={arXiv Preprint arXiv:2003.12408},
  year={2020}
}

@book{krishnamoorthy2009statistical,
  title={Statistical tolerance regions: theory, applications, and computation},
  author={Krishnamoorthy, Kalimuthu and Mathew, Thomas},
  year={2009},
  publisher={John Wiley \& Sons}
}

@misc{chernozhukov2018double,
  title={Double/debiased machine learning for treatment and structural parameters},
  author={Chernozhukov, Victor and Chetverikov, Denis and Demirer, Mert and Duflo, Esther and Hansen, Christian and Newey, Whitney and Robins, James},
  year={2018},
  publisher={Oxford University Press Oxford, UK}
}

@article{kallus2024localized,
  title={Localized debiased machine learning: Efficient inference on quantile treatment effects and beyond},
  author={Kallus, Nathan and Mao, Xiaojie and Uehara, Masatoshi},
  journal={Journal of Machine Learning Research},
  volume={25},
  pages={1--59},
  year={2024}
}

@article{lei2014distribution,
  title={Distribution-free prediction bands for non-parametric regression},
  author={Lei, Jing and Wasserman, Larry},
  journal={Journal of the Royal Statistical Society Series B: Statistical Methodology},
  volume={76},
  pages={71--96},
  year={2014},
  publisher={Oxford University Press}
}

@article{cheng2021robust,
  title={Robust and efficient semi-supervised estimation of average treatment effects with application to electronic health records data},
  author={Cheng, David and Ananthakrishnan, Ashwin N and Cai, Tianxi},
  journal={Biometrics},
  volume={77},
  pages={413--423},
  year={2021},
  publisher={Oxford University Press}
}

@article{prentice1989surrogate,
  title={Surrogate endpoints in clinical trials: definition and operational criteria},
  author={Prentice, Ross L},
  journal={Statistics in Medicine},
  volume={8},
  pages={431--440},
  year={1989},
  publisher={Wiley Online Library}
}

@article{frangakis2002principal,
  title={Principal stratification in causal inference},
  author={Frangakis, Constantine E and Rubin, Donald B},
  journal={Biometrics},
  volume={58},
  pages={21--29},
  year={2002},
  publisher={Oxford University Press}
}

@article{robins1992identifiability,
  title={Identifiability and exchangeability for direct and indirect effects},
  author={Robins, James M and Greenland, Sander},
  journal={Epidemiology},
  volume={3},
  pages={143--155},
  year={1992},
  publisher={LWW}
}

@article{gilbert2008evaluating,
  title={Evaluating candidate principal surrogate endpoints},
  author={Gilbert, Peter B and Hudgens, Michael G},
  journal={Biometrics},
  volume={64},
  pages={1146--1154},
  year={2008},
  publisher={Oxford University Press}
}

@article{freedman1992statistical,
  title={Statistical validation of intermediate endpoints for chronic diseases},
  author={Freedman, Laurence S and Graubard, Barry I and Schatzkin, Arthur},
  journal={Statistics in Medicine},
  volume={11},
  pages={167--178},
  year={1992},
  publisher={Wiley Online Library}
}

@article{fleming1996surrogate,
  title={Surrogate end points in clinical trials: are we being misled?},
  author={Fleming, Thomas R and DeMets, David L},
  journal={Annals of Internal Medicine},
  volume={125},
  pages={605--613},
  year={1996},
  publisher={American College of Physicians}
}

@article{athey2020combining,
  title={Combining experimental and observational data to estimate treatment effects on long term outcomes},
  author={Athey, Susan and Chetty, Raj and Imbens, Guido},
  journal={arXiv Preprint arXiv:2006.09676},
  year={2020}
}

@article{parast2016robust,
  title={Robust estimation of the proportion of treatment effect explained by surrogate marker information},
  author={Parast, Layla and McDermott, Mary M and Tian, Lu},
  journal={Statistics in Medicine},
  volume={35},
  pages={1637--1653},
  year={2016},
  publisher={Wiley Online Library}
}

@article{wang2020model,
  title={Model-free approach to quantifying the proportion of treatment effect explained by a surrogate marker},
  author={Wang, Xuan and Parast, Layla and Tian, LU and Cai, Tianxi},
  journal={Biometrika},
  volume={107},
  pages={107--122},
  year={2020},
  publisher={Oxford University Press}
}

@article{price2018estimation,
  title={Estimation of the optimal surrogate based on a randomized trial},
  author={Price, Brenda L and Gilbert, Peter B and van der Laan, Mark J},
  journal={Biometrics},
  volume={74},
  pages={1271--1281},
  year={2018},
  publisher={Wiley Online Library}
}

@article{han2022identifying,
  title={Identifying surrogate markers in real-world comparative effectiveness research},
  author={Han, Larry and Wang, Xuan and Cai, Tianxi},
  journal={Statistics in Medicine},
  volume={41},
  pages={5290--5304},
  year={2022},
  publisher={Wiley Online Library}
}

@article{wang2023robust,
  title={Robust approach to combining multiple markers to improve surrogacy},
  author={Wang, Xuan and Parast, Layla and Han, Larry and Tian, Lu and Cai, Tianxi},
  journal={Biometrics},
  volume={79},
  pages={788--798},
  year={2023},
  publisher={Wiley Online Library}
}

@article{buyse2000validation,
  title={The validation of surrogate endpoints in meta-analyses of randomized experiments},
  author={Buyse, Marc and Molenberghs, Geert and Burzykowski, Tomasz and Renard, Didier and Geys, Helena},
  journal={Biostatistics},
  volume={1},
  pages={49--67},
  year={2000},
  publisher={Oxford University Press}
}

@book{burzykowski2005evaluation,
  title={The evaluation of surrogate endpoints},
  author={Burzykowski, Tomasz and Buyse, Marc and Molenberghs, Geert},
  volume={427},
  year={2005},
  publisher={Springer}
}

@article{alonso2016information,
  title={An information-theoretic approach for the evaluation of surrogate endpoints based on causal inference},
  author={Alonso, Ariel and Van der Elst, Wim and Molenberghs, Geert and Buyse, Marc and Burzykowski, Tomasz},
  journal={Biometrics},
  volume={72},
  pages={669--677},
  year={2016},
  publisher={Wiley Online Library}
}

@article{elliott2023surrogate,
  title={Surrogate endpoints in clinical trials},
  author={Elliott, Michael R},
  journal={Annual Review of Statistics and its Application},
  volume={10},
  pages={75--96},
  year={2023},
  publisher={Annual Reviews}
}

@article{gilbert2024surrogate,
  title={A Surrogate Endpoint Based Provisional Approval Causal Roadmap},
  author={Gilbert, Peter B and Peng, James and Han, Larry and Lange, Theis and Lu, Yun and Nie, Lei and Shih, Mei-Chiung and Waddy, Salina P and Wiley, Ken and Yann, Margot and others},
  journal={arXiv Preprint arXiv:2407.06350},
  year={2024}
}

@article{sesia2020comparison,
  title={A comparison of some conformal quantile regression methods},
  author={Sesia, Matteo and Cand{\`e}s, Emmanuel J},
  journal={Stat},
  volume={9},
  pages={e261},
  year={2020},
  publisher={Wiley Online Library}
}

@article{kuchibhotla2023nested,
  title={Nested conformal prediction sets for classification with applications to probation data},
  author={Kuchibhotla, Arun K and Berk, Richard A},
  journal={The Annals of Applied Statistics},
  volume={17},
  pages={761--785},
  year={2023},
  publisher={Institute of Mathematical Statistics}
}

@book{imbens2015causal,
  title={Causal inference in statistics, social, and biomedical sciences},
  author={Imbens, Guido W and Rubin, Donald B},
  year={2015},
  publisher={Cambridge University Press}
}

@book{van2000asymptotic,
  title={Asymptotic Statistics},
  author={Van der Vaart, Aad W},
  volume={3},
  year={2000},
  publisher={Cambridge University Press}
}

@article{gilbert2022covid,
  title={A Covid-19 milestone attained—a correlate of protection for vaccines},
  author={Gilbert, Peter B and Donis, Ruben O and Koup, Richard A and Fong, Youyi and Plotkin, Stanley A and Follmann, Dean},
  journal={New England Journal of Medicine},
  volume={387},
  pages={2203--2206},
  year={2022},
  publisher={Mass Medical Soc}
}

@article{krause2022making,
  title={Making more {COVID-19} vaccines available to address global needs: {Considerations} and a framework for their evaluation},
  author={Krause, Philip R and Arora, Narendra and Dowling, William and Mu{\~n}oz-Fontela, C{\'e}sar and Funnell, Simon and Gaspar, Rogerio and Gruber, Marion F and Hacker, Adam and Henao-Restrepo, Ana Maria and Plotkin, Stanley and others},
  journal={Vaccine},
  volume={40},
  pages={5749},
  year={2022},
  publisher={Elsevier}
}

@book{little2019statistical,
  title={Statistical analysis with missing data},
  author={Little, Roderick JA and Rubin, Donald B},
  volume={793},
  year={2019},
  publisher={John Wiley \& Sons}
}

@book{wasserman2013all,
  title={All of statistics: a concise course in statistical inference},
  author={Wasserman, Larry},
  year={2013},
  publisher={Springer Science \& Business Media}
}

@article{chaisinanunkul2015adopting,
  title={Adopting a patient-centered approach to primary outcome analysis of acute stroke trials using a utility-weighted modified Rankin scale},
  author={Chaisinanunkul, Napasri and Adeoye, Opeolu and Lewis, Roger J and Grotta, James C and Broderick, Joseph and Jovin, Tudor G and Nogueira, Raul G and Elm, Jordan J and Graves, Todd and Berry, Scott and others},
  journal={Stroke},
  volume={46},
  pages={2238--2243},
  year={2015},
  publisher={Am Heart Assoc}
}

@article{nogueira2018thrombectomy,
  title={Thrombectomy 6 to 24 hours after stroke with a mismatch between deficit and infarct},
  author={Nogueira, Raul G and Jadhav, Ashutosh P and Haussen, Diogo C and Bonafe, Alain and Budzik, Ronald F and Bhuva, Parita and Yavagal, Dileep R and Ribo, Marc and Cognard, Christophe and Hanel, Ricardo A and others},
  journal={New England Journal of Medicine},
  volume={378},
  pages={11--21},
  year={2018},
  publisher={Mass Medical Soc}
}

@article{imbens2024long,
  title={Long-term causal inference under persistent confounding via data combination},
  author={Imbens, Guido and Kallus, Nathan and Mao, Xiaojie and Wang, Yuhao},
  journal={Journal of the Royal Statistical Society Series B: Statistical Methodology},
  pages={qkae095},
  year={2024},
  publisher={Oxford University Press UK}
}
\bibliographystyle{dcu}

\newpage

\appendix
\counterwithin{theorem}{section}
\counterwithin{lemma}{section}
\counterwithin{corollary}{section}
\counterwithin{table}{section}
\section*{Supplementary Materials}

The Supplementary Materials include the following information: Section \ref{sec:proofs} provides the proofs of Theorems \ref{thm:EIF1} to \ref{thm:EIF-theta_C}, as well as of Corollaries \ref{coro:v1_reduction} and \ref{coro:v1_reduction_D0}. We provide additional simulation results in Section \ref{more:sims}, and additional results for real-data analysis in Section \ref{more:real-data}.


\section{Technical Proofs and Details}\label{sec:proofs}

\subsection{Proof of Theorem \ref{thm:EIF1}}

   Let $t$ denote the index for the parametric submodels $f_t(\mathcal{O})$ of the observed data with true density function evaluated at $t=t^*$, i.e., $f(\mathcal{O})=f_{t^*}(\mathcal{O})$, where 
   \begin{align*}
       f(\mathcal{O})& = 
   f(X)\times e_A(X)^A\{1-e_A(X)\}^{1-A} \times 
   e_D(X, A)^D\{1-e_D(X, A)\}^{1-D}\\
   &\times f(S\mid D, A, X)
   \times f(Y\mid S, D, A, X),
   \end{align*}
    the observed score function is derived by the pathwise derivatives of $\log f_t(\mathcal{O})$ with respect to $t$, given by $s_t(\mathcal{O})=s_t(Y\mid S, D, A, X)+s_t(S\mid D, A, X) + s_t(D\mid A,X) +  s_t(A\mid X) + s_t(X)$. The tangent space corresponding to this model is
   $\Lambda = \Lambda_X \oplus \Lambda_{A\mid X} \oplus \Lambda_{D\mid A, X} \oplus \Lambda_{S\mid D, A, X} \oplus \Lambda_{Y\mid S, D, A, X}$, where $\Lambda_X$, $\Lambda_{A\mid X}$, $\Lambda_{D\mid A, X}$, $\Lambda_{S\mid D, A, X}$, and $\Lambda_{Y\mid S, D, A, X}$ are the mean square closures of the score vector of the submodels:
    \begin{align*}
        &\Lambda_X = \{\Gamma(X): \int \Gamma(x)f(x)dx=0\},\\
        &\Lambda_{A\mid X} = \{
        \{A-e_A(X)\}a(X)
        \},\quad \Lambda_{D\mid A, X} = \{A\{D-e_D(X, A)\}b(X)\},\\
        &\Lambda_{S\mid D, A, X} = \{\Gamma(S, D, A, X): \int \Gamma(s, D, A, X)f(S\mid D, A, X)ds=0\},\\
        &\Lambda_{Y\mid S, D, A, X} = \{\Gamma(Y, S, D, A, X): \int \Gamma(y, S, D, A, X)f(y \mid S, D, A, X)dy=0\},
    \end{align*}
    for any two arbitrary square-integrable measurable functions $a(X)$ and $b(X)$.
    
      Let  $u(R_1,r_{\alpha,1})=\mathbf{1}(R_1\leq r_{\alpha,1}\mid A=0, D=1) -(1-\alpha)$, and $r_1(t)$ be the corresponding $(1-\alpha)$-quantile for $R_1\mid A=0,D=1$ under the parametric submodels, which satisfies $E_t[u\{R_1,r_1(t)\}]=0$ with $r(t^*)=r_{\alpha,1}$. Next, we can show that 
   \begin{align*}
       &0 = \frac{\partial }{\partial t}E_t[u\{R_1,r_1(t)\}]\big|_{t=t^*} =
       \frac{\partial }{\partial t}E_t[u\{R_1,r_1(t^*)\}]\big|_{t=t^*}+ 
       \frac{\partial }{\partial r_1}E_{t^*}[u\{R_1,r_1(t^*)\}]
       \frac{\partial r_1(t)}{\partial t}\big|_{t=t^*},
   \end{align*}
   which leads to
   \begin{align*}
      &\frac{\partial r_1(t)}{\partial t}\big|_{t=t^*} = 
      \left[-\frac{\partial }{\partial r_1}E_{t^*}\{u(R_1,r_{\alpha, 1})\}\right]^{-1} \frac{\partial }{\partial t}E_t\{u(R_1,r_{\alpha,1})\}\big|_{t=t^*}\propto \frac{\partial }{\partial t}E_t\{u(R_1,r_{\alpha,1})\}\big|_{t=t^*}.
\end{align*}
Under Assumptions \ref{assump:overlap} to \ref{assump:unconfoundedness}, we have
\begin{align*}
      &\frac{\partial }{\partial t}E_t\{u(R_1,r_{\alpha,1})\}\big|_{t=t^*} = 
      \frac{\partial}{\partial t}P_t(R_1<{r}_{\alpha,1}\mid A = 0, D = 1)\big|_{t=t^*}\\
      &\propto 
      \frac{\partial}{\partial t}E_{X,t}[E_{S,t}\{P_{Y,t}(R_1<{r}_{\alpha,1}\mid A = 1, D = 1, S, X)\mid A=1, X\}\mid A=0, D=1]\big|_{t=t^*}\\
      &\propto 
      \frac{\partial}{\partial t}E_{X,t}[(1-A)DE_{S,t}\{P_{Y,t}(R_1<{r}_{\alpha,1}\mid A = 1, D = 1, S, X)\mid A=1, X\}]\big|_{t=t^*}\\
      &\propto 
      \frac{\partial}{\partial t}E_{X,t}[
      P_t(A=0, D=1\mid X)
      E_{S,t}\{P_{Y,t}(R_1<{r}_{\alpha,1}\mid A = 1, D = 1, S, X)\mid A=1, X\}]\big|_{t=t^*}\\
      & = T_1 + T_2 + T_3+ T_4,
   \end{align*} 
   which is constituted by five terms $T_1$, $T_2$, $T_3$, and $T_4$. We analyze these four terms separately,
    \begin{align*}
        T_1 &=  
        \frac{\partial}{\partial t}E_{X,t}\{
      P(A=0, D=1\mid X)
      m_1(r,X)\}\big|_{t=t^*}\\
      &=E_X[
      P(A=0, D=1\mid X)\{m_1(r,X)-(1-\alpha)\}s(X)],\\
              T_2 &= \frac{\partial}{\partial t}E_{X}\{
      P_t(A=0, D=1\mid X)
      m_1(r,X)\}\big|_{t=t^*}\\
      &=E
      \{[(1-A)D - \{1-e_A(X)\}e_D(X,0)]m_1(r,X)s(D,A\mid X)\},\\
              T_3 & = \frac{\partial}{\partial t}E_{X}[
      P(A=0, D=1\mid X)
    E_{S,t}\{\tilde{m}_1(r,X,S)\mid A=1, X\}]\big|_{t=t^*}\\
    &= E\left[ P(A=0,D=1\mid X) \frac{A}{e_A(X)}\left\{\Tilde{m}_1(r,X,S) - m_1(r,X)\right\}
    s(S\mid D, A, X)\right],
    \end{align*}
    and
    \begin{align*}
        T_4 & = \frac{\partial}{\partial t}E_{X}[
      P(A=0, D=1\mid X)
      E_{S}\{P_{Y,t}(R_1<{r}_{\alpha,1}\mid A = 1, D = 1, S, X)\mid A=1, X\}]\big|_{t=t^*} \\
      & = \frac{\partial}{\partial t}E_{X}\left(
      P(A=0, D=1\mid X)
      E\left[\frac{D\left\{\mathbf{1}(R_1\leq r_{\alpha,1}) -\Tilde{m}_1(r,X,S)\right\}}{e_D(X,1)} s(Y\mid S, D, A, X)\mid A=1,X\right]\right) \\
      & = \frac{\partial}{\partial t}E\left[
      P(A=0, D=1\mid X)
      \frac{AD\left\{\mathbf{1}(R_1\leq r_{\alpha,1}) -\Tilde{m}_1(r,X,S)\right\}}{e_A(X)e_D(X,1)} s(Y\mid S, D, A, X)\right].
    \end{align*}
    Putting these terms together, we can show that 
    $$
    \frac{\partial r(t)}{\partial t}\big|_{t=t^*}
    \propto \frac{\partial }{\partial t}E_t\{u(R_1,r_{\alpha,1})\}\big|_{t=t^*} = 
E\{\psi_1(r_{\alpha,1}, W;m,\Tilde{m},e_D,\pi_A)s(Y,S,D,A,X)\},
    $$
and
\begin{align*}
&\psi_1(r_{\alpha,1},W;m,\Tilde{m},e_D,\pi_A) = 
    D(1-A)\{m_1(r_{\alpha,1}, X) - (1 - \alpha)\} \\
    & + A\pi_A(X) e_D(X,0)\{\Tilde{m}_1(r_{\alpha,1},X,S) - m_1(r_{\alpha,1},X)\}  \\
    & + \frac{AD\pi_A(X)e_D(X,0)}{e_D(X,1)}
    \{\mathbf{1}(R_1<{r}_{\alpha,1}) - \Tilde{m}_1(r_{\alpha,1},X,S)\},
\end{align*}
where 
\begin{align*}
    & D(1-A)\{m_1(r_{\alpha,1}, X) - (1 - \alpha)\} \in \Lambda_X\\
    & A\pi_A(X) e_D(X,0)\{\Tilde{m}_1(r_{\alpha,1},X,S) - m_1(r_{\alpha,1},X)\} \in \Lambda_{S\mid D, A, X}\\
    & \frac{AD\pi_A(X)e_D(X,0)}{e_D(X,1)}
    \{\mathbf{1}(R_1<{r}_{\alpha,1}) - \Tilde{m}_1(r_{\alpha,1},X,S)\} \in \Lambda_{Y\mid S, D, A, X}.
\end{align*}
Thus, it completes the proof of Theorem \ref{thm:EIF1} by the definition of efficient influence function.

\subsection{Proof of Theorem \ref{thm:DR_R}}

Under Lemma \ref{lem:Dr}, we can show that for the empirically estimated $\widehat{r}_\alpha$ using the second data fold $\mathcal{I}_2$ under Setting 2,
\begin{align*}
    &\mathbb{P}(R_1\leq \widehat{r}^{(S2)}_{\alpha,1} \mid A=0, D=1) - (1-\alpha) = 
    \frac{\mathbb{P}_{\mathcal{I}_2}\{\widehat{\psi}^{(S2)}_1(r_{\alpha,1},W)\}}{P(A=0, D=1)}\\
    &+ 
    \frac{\mathbb{P}\{\widehat{\psi}^{(S2)}_1(r_{\alpha,1},W)\} - \mathbb{P}_{\mathcal{I}_2}\{\widehat{\psi}^{(S2)}_1(r_{\alpha,1},W)\}}{P(A=0, D=1)} +\frac{\mathbb{P}\{\widehat{\psi}^{(S2)}_1(r_{\alpha,1},W) - 
    {\psi}^{(S2)}_1(r_{\alpha,1},W)\}}{P(A=0, D=1)}\\
    &\geq 0 + I_1 + I_2.
\end{align*}
Here, we use the fact that $\mathbb{P}_{\mathcal{I}_2}\{\widehat{\psi}^{(S2)}_1(r_{\alpha,1},W)\}\geq 0$ by definition. The term $I_1$ is negligible if $\psi_1^{(S2)}(r_{\alpha, 1}, W)$ belongs to a Donsker class \citep{van2000asymptotic}.  Even if the Donsker condition is not met, the sample-splitting procedure in \cite{chernozhukov2018double} can be used to assure that $I_1$ is negligible, where the first data fold $\mathcal{I}_1$ is used to estimate $\widehat{\pi}_A$, $\widehat{e}_D$, $\widehat{m}_a$ and $\widehat{\tilde{m}}$, and the second data fold $\mathcal{I}_2$ is used to compute $r_{\alpha, 1}$; see Lemma \ref{lem:empirical} for more details on the bound of $I_1$.

The term $I_2$ is the second-order remainder term, which is bounded by the product of estimation error for the nuisance functions:
\begin{align*}
    & \mathbb{P}\{\widehat{\psi}^{(S2)}_1(r,W) - 
    {\psi}^{(S2)}_1(r,W)\}  = 
\mathbb{P}\left[P(A=0\mid X)e_D(X,0)\{\widehat{m}_1(r, X) - m_1(r, X)\}\right] \\
& + \mathbb{P}\left[P(A=1\mid X) \widehat{\pi}_A(X)\widehat{e}_D(X,0) \{\widehat{\Tilde{m}}_1(r,X,S) - \widehat{m}_1(r,X)\}\right] \\ 
& + \mathbb{P}\left[P(A=1\mid X)\widehat{\pi}_A(X)\frac{e_D(X, 1)}{\widehat{e}_D(X, 1)}\widehat{e}_D(X, 0)\{{\Tilde{m}}_1(r,X,S) - \widehat{\Tilde{m}}_1(r,X,S)\}\right]\\
&=\mathbb{P}\left[P(A=0\mid X)e_D(X,0)\{\widehat{m}_1(r, X) - m_1(r_{\alpha,1}, X)\}\right] \\
& + \mathbb{P}\left[P(A=1\mid X) \widehat{\pi}_A(X)\widehat{e}_D(X,0) \{\widehat{\Tilde{m}}_1(r,X,S) - 
{\Tilde{m}}_1(r,X,S) + {{m}}_1(r,X) -\widehat{m}_1(r,X)\}\right] \\ 
& + \mathbb{P}\left[P(A=1\mid X)\widehat{\pi}_A(X)\frac{e_D(X, 1)}{\widehat{e}_D(X, 1)}\widehat{e}_D(X, 0)\{{\Tilde{m}}_1(r,X,S) - \widehat{\Tilde{m}}_1(r,X,S)\}\right]\\
&=
\mathbb{P}\left[ P(A=0\mid X)\left\{
\frac{\pi_A(X)e_D(X,0) - 
\widehat{\pi}_A(X)\widehat{e}_D(X,0)}{\pi_A(X)}
\right\}\{\widehat{m}_1(r, X) - m_1(r, X)\}\right] \\ 
&+ \mathbb{P}\left[P(A=1\mid X) \widehat{\pi}_A(X) \widehat{e}_D(X, 0)\left\{ 
\frac{\widehat{e}_D(X,1) -
e_D(X,1)}{\widehat{e}_D(X,1)}
\right\} \{\widehat{\Tilde{m}}_1(r,X,S) - 
{\Tilde{m}}_1(r,X,S)\}\right]\\
&\lesssim \|\widehat{\pi}_A(X) - {\pi}_A(X)\|\cdot \sup_r\|\widehat{m}_1(r, X) - m_1(r, X)\|\\
& + \|\widehat{e}_D(X,0) - {e}_D(X,0)\|\cdot \sup_r\|\widehat{m}_1(r, X) - m_1(r, X)\|\\
& + \|\widehat{e}_D(X,1) - {e}_D(X,1)\|\cdot \sup_r\|\widehat{\Tilde{m}}_1(r,X,S) - 
{\Tilde{m}}_1(r,X,S)\|,
\end{align*}
where the last inequality follows from the Cauchy–Schwarz inequality. Combining the bounds for $I_1$ and $I_2$ gives us
\begin{align*}
   &P(
R_a \leq \widehat{r}_{\alpha,a}^{(S2)}\mid A_i=1-a, D_i=1) \geq (1-\alpha) \\
& - C_0\frac{\pi_{0}\overline{e}_{0}(\underline{e}_{0}^{-1}+ m_{0} +\underline{e}_{0}^{-1}\tilde{m}_{0})}{P(A=1-a, D=1)}\sqrt{\frac{\log(1/\delta) + 1}{|\mathcal{I}_2|}}\\
   &-C_1\left\{\frac{\|\widehat{\pi}_A(X) - {\pi}_A(X)\|}{P(A=1-a, D=1)}\cdot \sup_r\|\widehat{m}_a(r,X) - m_a(r,X)\|\right.\\
   & \left.+ \frac{\|\widehat{e}_D(X,A) - e_D(X,A)\|}{P(A=1-a, D=1)}\cdot \sup_r\|\widehat{m}_a(r,X) - m_a(r,X)\|\right.\\
& \left.+ \frac{\|\widehat{e}_D(X,A) - e_D(X,A)\|}{P(A=1-a, D=1)}\cdot \sup_r\|\widehat{\Tilde{m}}_a(r, X,S) - 
{\Tilde{m}}_a(r,X,S)\|\right\},
\end{align*}
and the proof for $P(
R_a \leq \widehat{r}_{\alpha,a}^{(S1)}\mid A_i=1-a, D_i=1)$ under Settings 1 or 3 can be completed in a similar manner.

\subsection{Proof of Corollary \ref{coro:v1_reduction}}
By straightforward algebra, the semi-parametric lower bound for estimating $r_{\alpha, 1}$ without surrogates for $\psi_1^{(S1)}(r_{\alpha,1}, W;m,\Tilde{m},e_D,\pi_A)$ in Theorem \ref{thm:EIF1} is
    \begin{align*}
        V^{(S1)} & = 
    \var\{\psi_1^{(S1)}(r_{\alpha,1}, W;m,\Tilde{m},e_D,\pi_A)\} =
    V_1 + V_2 + V_3 +2V_{12} - 2V_{13} - 2V_{23},
    \end{align*}
    where
    \begin{align*}
        &V_1 = \var[D(1-A)\{m_1(r_{\alpha,1}, X) - (1 - \alpha)\}],\\
        &V_2 = 
        \var \left\{\frac{AD\pi_A(X)e_D(X,0)}{e_D(X,1)} \mathbf{1}(R_1<{r}_{\alpha,1})\right\},
         \\
    & V_3 = \var \left\{\frac{AD\pi_A(X)e_D(X,0)}{e_D(X,1)}
    m_1(r_{\alpha,1},X) \right\},\\
    &V_{12} = V_{13} \\
    & = \cov \left[ D(1-A)\{m_1(r_{\alpha,1}, X) - (1 - \alpha)\}, \frac{AD\pi_A(X)e_D(X,0)}{e_D(X,1)} m_1(r_{\alpha,1},X)\right],\\
    &V_{23} = V_3 \\
    &=\cov \left[ \frac{AD\pi_A(X)e_D(X,0)}{e_D(X,1)} \mathbf{1}(R_1<{r}_{\alpha,1}), \frac{AD\pi_A(X)e_D(X,0)}{e_D(X,1)} m_1(r_{\alpha,1},X)\right],\\
    &=\var \left\{\frac{AD\pi_A(X)e_D(X,0)}{e_D(X,1)}
    m_1(r_{\alpha,1},X) \right\}.
    \end{align*}
    In addition, we can show that the efficiency lower bound with surrogates is
    \begin{align*}
    &V^{(S2)} = 
    \var\{\psi_1^{(S2)}(W;m,\Tilde{m},e_D,\pi_A)\} \\
    & = \Tilde{V}_1 + \Tilde{V}_2 + \Tilde{V}_3 + \Tilde{V}_4 + 
    2\Tilde{V}_{12} + 2\Tilde{V}_{13} - 2\Tilde{V}_{14} + 2\Tilde{V}_{23} - 2\Tilde{V}_{24} - 2\Tilde{V}_{34},
    \end{align*}
    where
    \begin{align*}
    &\Tilde{V}_1 = \var[D(1-A)\{m_1(r_{\alpha,1}, X) - (1 - \alpha)\}]
         \\
    & \Tilde{V}_2 = \var[A\pi_A(X) e_D(X,0)\{\Tilde{m}_1(r_{\alpha,1},X,S) - m_1(r_{\alpha,1},X)\}], \\
    & \Tilde{V}_3 = \var\left\{
    \frac{AD\pi_A(X)e_D(X,0)}{e_D(X,1)}\mathbf{1}(R_1<{r}_{\alpha,1})
    \right\}, \\
    & \Tilde{V}_4 = \var\left\{
    \frac{AD\pi_A(X)e_D(X,0)}{e_D(X,1)} \Tilde{m}_1(r_{\alpha,1},X,S)\right\},\\
    &\Tilde{V}_{12} =
    \cov \left[D(1-A)\{m_1(r_{\alpha,1}, X) - (1 - \alpha)\},
    A\pi_A(X) e_D(X,0)\{\Tilde{m}_1(r_{\alpha,1},X,S) - m_1(r_{\alpha,1},X)\}\right] = 0, \\
    &\Tilde{V}_{13} = \Tilde{V}_{14},\quad \Tilde{V}_{23} = \Tilde{V}_{24}, \quad \Tilde{V}_{34}=\Tilde{V}_4.
    \end{align*}
    Hence, we can verify that under Assumptions \ref{assump:MAR} and \ref{assump:unconfoundedness}, 
    \begin{align*}
        & V^{(S1)} - V^{(S2)} = \Tilde{V}_4 - V_3 - \Tilde{V}_2\\
        & = \var\left\{
    \frac{AD\pi_A(X)e_D(X,0)}{e_D(X,1)} \Tilde{m}_1(r_{\alpha,1},X,S)\right\} - \var \left\{\frac{AD\pi_A(X)e_D(X,0)}{e_D(X,1)}
    m_1(r_{\alpha,1},X) \right\}\\
    &-\var[A\pi_A(X) e_D(X,0)\{\Tilde{m}_1(r_{\alpha,1},X,S) - m_1(r_{\alpha,1},X)\}]\\
    & = E\left[
    \frac{P(A=1,D=1\mid X)\pi_A^2(X)e_D^2(X,0)}{e_D^2(X,1)} \var\left\{\Tilde{m}_1(r_{\alpha,1},X,S)\mid X, A=1, D=1\right\}
    \right]\\
    &-E[e_A(X)\pi_A^2(X) e_D^2(X,0)\var\{\Tilde{m}_1(r_{\alpha,1},X,S)\mid X, A=1\}]\\
        & = 
        E \left[
        \frac{1 - e_D(X,1)}{e_D(X,1)}
        \frac{e_D^2(X,0)\{1-e_A(X)\}^2}{e_A(X)}
        \var \{\Tilde{m}_1(r_{\alpha,1},X,S)\mid X\}
        \right],
    \end{align*}
    since $\Tilde{V}_1 = V_1$ and $\Tilde{V}_3 = V_2$, which completes the proof of Corollary \ref{coro:v1_reduction}.

\subsection{Proof of Theorem \ref{thm:EIF-theta_C}}

  Under Assumptions \ref{assump:overlap} to \ref{assump:unconfoundedness}, the $(1-\alpha)$-th quantile $r_{\gamma,C}$ of $R_C\mid D=0$ can be identified by
\begin{align*}
    &P\{Y(1)-Y(0)\in C\mid D=0\} = E[P\{Y(1)-Y(0)\in C\mid X, D=0\}\mid D=0]\\
    & = E[P\{Y(1)-Y(0)\in C\mid X, D=1\}\mid D=0] \\
    & =
    E[P\{R_C < r_{\gamma,C}\mid X, D=1\}\mid D = 0]\\
    & = \frac{1}{P(D=0)} E[(1-D)P\{R_C < r_{\gamma,C}\mid X, D=1\}]\\
    & =
    \frac{1}{P(D=0)}
    E[P(D=0\mid X)P\{R_C < r_{\gamma,C}\mid X, D=1\}]\\
    &=
    \frac{1}{P(D=0)}
    E(P(D=0\mid X)E[P\{R_C < r_{\gamma,C}\mid X, S, D=1\}\mid X]),
\end{align*}
where the last equation holds as $D\perp \{Y(a),S(a)\}\mid X$ by Assumption \ref{assump:MAR}. Next, we can derive the EIF for $r_{\gamma,C}$, similar to the proof of Theorem \ref{thm:EIF1}:
\begin{align*}
    &\psi_C(W; m_C, \Tilde{m}_C,e_D, \pi_A) = 
    \{1 - e_D(X)\}m_C(r_{\gamma,C}, X) +
    \{1 - D - 1 + e_D(X)\}m_C(r_{\gamma,C}, X)\\
    &+\{1-e_D(X)\}\{\Tilde{m}_C(r_{\gamma,C}, X,S) - m_C(r_{\gamma,C}, X)\} \\
    & + \{1-e_D(X)\}\frac{D}{e_D(X)}\{\mathbf{1}(R_C<r_{\gamma,C}) - \Tilde{m}_C(r_{\gamma,C}, X,S)\}\\
    & = (1-D)m_C(r_{\gamma,C}, X) \\
    &+\{1-e_D(X)\}\{\Tilde{m}_C(r_{\gamma,C}, X,S) - m_C(r_{\gamma,C}, X)\} + 
    De_D(X)\{\mathbf{1}(R_C<r_{\gamma,C}) - \Tilde{m}_C(r_{\gamma,C}, X,S)\},
\end{align*}
where $\Tilde{m}_C(r_{\gamma,C}, X,S) =P(R_C <r_{\gamma,C} \mid X, S, D=1)$, $m_C(r_{\gamma,C}, X) = E\{\Tilde{m}_C(r_{\gamma,C}, X,S)\mid X\}$, and $e_D(X)=\{1-e_D(X)\}/e_D(X)$.

\subsection{Proof of Corollary \ref{coro:v1_reduction_D0}}
The semi-parametric efficiency lower bound for estimating $r_{\gamma,C}$ without surrogates is
\begin{align*}
    V_C^{(S1)} &= \var\{\psi_C^{(S1)}(r_{\gamma,C},W; m_C,e_D, \pi_A)\} \\
    &= V_1^C + V_2^C + V_3^C + 2V_{12}^C - 2V_{13}^C - 2V_{23}^C,
\end{align*}
where 
\begin{align*}
    &V_1^C = \var\left[
    (1-D)\{m_C(r_{\gamma,C}, X) - (1-\gamma)\}
    \right],\\
    &V_2^C = \var\left[
    D\pi_D(X)\mathbf{1}(R_C<r_{\gamma,C})
    \right], \quad V_3^C = \var\left\{
    D\pi_D(X)m_C(r_{\gamma,C}, X)
    \right\},\\
    &V_{12}^C = V_{13}^C = 0,\\
    &V_{23}^C = \cov \left[  D\pi_D(X)\mathbf{1}(R_C<r_{\gamma,C}), D\pi_D(X)m_C(r_{\gamma,C}, X)\right] = V_3^C.
\end{align*}
Similarly, we can show that the efficiency lower bound with surrogates is
\begin{align*}
    \tilde{V}_C^{(S2)} &= \var\{\psi_C^{(S2)}(r_{\gamma,C},W; m_C,\tilde{m}_C, e_D, \pi_A)\} \\
    &= \tilde{V}_1^C + \tilde{V}_2^C + \tilde{V}_3^C + \tilde{V}_4^C \\
    & + 2\tilde{V}_{12}^C + 2\tilde{V}_{13}^C - 2\tilde{V}_{14}^C+ 2\tilde{V}_{23}^C -2\tilde{V}_{24}^C -2\tilde{V}_{34}^C
\end{align*}
where 
\begin{align*}
&   \tilde{V}_1^C = \var \left[
(1-D)\{m_C(r_{\gamma,C},X)- (1-\gamma)\}
\right], \\
& \tilde{V}_2^C = \var\left[
\{1-e_D(X)\}\{\Tilde{m}_C(r_{\gamma,C},X,S) - m_C(r_{\gamma,C},X)\}
\right]\\
& \tilde{V}_3^C = \var\left\{
D\pi_D(X)\mathbf{1}(R_C<r_{\gamma,C})
\right\},\quad \tilde{V}_4^C = \var\left\{
D\pi_D(X)\Tilde{m}_C(r_{\gamma,C},X,S)
\right\},\\
& \tilde{V}_{12}^C =\tilde{V}_{13}^C = \tilde{V}_{14}^C = 0,\\
& \tilde{V}_{23}^C = \tilde{V}_{24}^C\\
& = \cov \left[  \{1-e_D(X)\}\{\Tilde{m}_C(r_{\gamma,C},X,S) - m_C(X)\}, D\pi_D(X)\Tilde{m}_C(r_{\gamma,C},X,S)\right],\\
& \tilde{V}_{34}^C = \cov \left[  D\pi_D(X)\mathbf{1}(R_C<r_{\gamma,C}), D\pi_D(X)\Tilde{m}_C(r_{\gamma,C},X,S)\right] = \tilde{V}_{4}^C.
\end{align*}
Therefore, we can verify that
\begin{align*}
    & V_C^{(S1)} - V_C^{(S2)} = \tilde{V}_{4}^C - V_3^C - \tilde{V}_{2}^C\\
    & = \var\left\{
D\pi_D(X)\Tilde{m}_C(r_{\gamma,C},X,S)
\right\} - \var\left\{
    D\pi_D(X)m_C(r_{\gamma,C},X)
    \right\} \\
    &- 
\var\left[
\{1-e_D(X)\}\{\Tilde{m}_C(r_{\gamma,C},X,S) - m_C(r_{\gamma,C},X)\}
\right]\\
& = E\left\{
D\pi_D^2(X)\Tilde{m}_C^2(r_{\gamma,C},X,S)
\right\} - 
E\left\{
    D\pi_D^2(X)m_C^2(r_{\gamma,C},X)
    \right\}\\&- 
E\left[
\{1-e_D(X)\}^2\{\Tilde{m}_C(r_{\gamma,C},X,S) - m_C(r_{\gamma,C},X)\}^2
\right]\\
& = E\left[
\frac{\{1-e_D(X)\}^2}{e_D(X)}\{\Tilde{m}_C^2(r_{\gamma,C},X,S) - m_C^2(r_{\gamma,C},X)\}
\right]\\
&- 
E\left[
\{1-e_D(X)\}^2\{\Tilde{m}_C^2(r_{\gamma,C},X,S) - m_C^2(r_{\gamma,C},X)\}
\right]\\
    & = E \left[
    \pi_D(X)\{1-e_D(X)\}^2
    \var \{\Tilde{m}_C(r_{\gamma,C},X,S)\mid X\}
    \right],
\end{align*}
where $V_1^C=\tilde{V}_1^C$ and $V_2^C=\tilde{V}_3^C$. Hence, it completes the proof of Corollary \ref{coro:v1_reduction_D0}.

\subsection{Additional technical details}
\label{subsec:thm_additional}

\subsubsection{Proof of Lemma \ref{lem:assumption}}
The proof of Lemma 2.2 in \cite{kallus2020role} shows that $D\perp \{Y(a), S(a)\}\mid X, A$ under Assumptions \ref{assump:external_validtiy} and \ref{assump:MAR}. Under Assumption \ref{assump:unconfoundedness}, we have $(A, D)\perp \{Y(a), S(a)\}\mid X$, which proves Condition (a), that is, $D\perp\{Y(a), S(a)\}\mid X$, and Conditions (b) and (c) $A\perp \{Y(a), S(a))\}\mid X, D$ by Theorem 17.2 in \cite{wasserman2013all}.


\subsubsection{Proof of Lemma \ref{lem:Dr}}
On the one hand, we can show that
\begin{align}
    &P(R_1\leq r_{\alpha,1} \mid A=0,D=1)
    = E\{E(R_1\leq r_{\alpha,1} \mid X, A=0, D=1)\mid A=0, D=1\} \nonumber\\
    &= E\{E(R_1\leq r_{\alpha,1} \mid X, A=1, D=1)\mid A=0, D=1\} \text{ (Assumption \ref{assump:unconfoundedness})} \nonumber\\
    & = \int_\mathcal{X}
    \int_{-\infty}^{r_{\alpha,1}}
    f(r_1\mid x, A=1,D=1)f(x\mid A=0, D=1) dr_1 dx
    \nonumber\\
    & = \int_\mathcal{X}
    \int_{-\infty}^{r_{\alpha,1}}
    \frac{f(x\mid A=0, D=1)}{f(x\mid A=1, D=1)}
    f(r_1\mid x, A=1,D=1)f(x\mid A=1, D=1) dr_1 dx
    \nonumber\\
    &=E \left\{\frac{f(X\mid A=0, D=1)}{f(X\mid A=1, D=1)}\mathbf{1}(R_1\leq r_{\alpha, 1})\mid A=1, D=1\right\}
    \label{eq:prove_cov1}.
\end{align}
On the other hand, we can prove that
\begin{align*}
   & E\{\psi_1(r_{\alpha,1}, W;m,\tilde{m},e_D,\pi_A)\} = E [D(1-A)\{m_1(r_{\alpha,1},X)- (1-\alpha)\}] \\
    & = E\left\{
    P(A=0, D=1\mid X)P(R_1\leq r_{\alpha,1}\mid X)\right\} -
     P(A=0, D=1)(1-\alpha)\\
     & = E\left\{AD \frac{P(A=0, D=1\mid X)}{P(A=1, D=1\mid X)} P(R_1\leq r_{\alpha,1}\mid X)\right\} -
     P(A=0, D=1)(1-\alpha)\\
     & = \frac{P(A=0, D=1)}{P(A=1, D=1)}
     E\left\{AD \frac{f(X\mid A=0, D=1)}{f(X\mid A=1, D=1)} P(R_1\leq r_{\alpha,1}\mid X)\right\}\\
     &-
     P(A=0, D=1)(1-\alpha)\\
    &= 
    P(A=0, D=1)
    E\left\{\frac{f(X\mid A=0, D=1)}{f(X\mid A=1, D=1)}\mathbf{1}(R_1\leq r_{\alpha,1})\mid A=1, D=1\right\} \\
    &- P(A=0, D=1)(1-\alpha).
\end{align*}
Combine it with (\ref{eq:prove_cov1}), we have $P(R_1\leq r_{\alpha,1} \mid A=0,D=1) = 1-\alpha + E\{\psi_1(r_{\alpha,1},W)\}/P(A=0, D=1)$. 

\subsubsection{Proof of Lemma \ref{lem:empirical}}
\begin{lemma}\label{lem:empirical}
    Under the regularity conditions (A1) and (A2), there exists some constant $C_0$ such that for any $\delta>0$, 
    $$
    P\left(|I_1|\leq \frac{C_0 \pi_{0}\overline{e}_{0}(\underline{e}_{0}^{-1}+ m_{0} +\underline{e}_{0}^{-1}\tilde{m}_{0})}{P(A=0,D=1)} \sqrt{\frac{\log(1/\delta) + 1}{N}}\mid \mathcal{I}_1\right)\geq 1-\delta.
    $$
\end{lemma}
The proof of Lemma \ref{lem:empirical} is adapted from Theorem 3 in \cite{yang2024doubly}.
Specifically, we expand the numerator of $I_1$ into the following four parts,
\begin{align}
    &\mathbb{P}\{\widehat{\psi}^{(S2)}_1(r_{\alpha,1},W)\} - \mathbb{P}_{\mathcal{I}_2}\{\widehat{\psi}^{(S2)}_1(r_{\alpha,1},W)\} \nonumber\\
    &=\left[\frac{1}{|\mathcal{I}_2|}\sum_{i\in \mathcal{I}_2}
    \frac{A_iD_i\widehat{\pi}_A(X_i)\widehat{e}_D(X_i,0)\mathbf{1}(R_1 < r_{\alpha, 1})}{\widehat{e}_D(X_i,1)} -
    \mathbb{P}
    \left\{\frac{AD\widehat{\pi}_A(X)\widehat{e}_D(X,0)}{\widehat{e}_D(X,1)} \mathbf{1}(R_1 < r_{\alpha, 1})\right\}\right]\label{eq:I_1:part1}\\
    &\begin{aligned}
 & + \frac{1}{|\mathcal{I}_2|}\sum_{i\in \mathcal{I}_2} \left\{D_i(1-A_i)- 
    A_i\widehat{\pi}_A(X_i) \widehat{e}_D(X_i,0)\right\}\widehat{m}_1(r_{\alpha,1},X_i)\\
& - \mathbb{P}\left\{D(1-A)- 
    A\widehat{\pi}_A(X) \widehat{e}_D(X,0)\right\}\widehat{m}_1(r_{\alpha,1},X) \\
\end{aligned}\label{eq:I_1:part2}\\
    &\begin{aligned}
 & + \frac{1}{|\mathcal{I}_2|}\sum_{i\in \mathcal{I}_2}
     \left\{A_i\pi_A(X_i) e_D(X_i,0) - \frac{A_iD_i\pi_A(X_i)e_D(X_i,0)}{e_D(X_i,1)}\right\}\widehat{\tilde{m}}_1(r_{\alpha,1},X_i, S_i)\\ 
     & - \mathbb{P}
     \left\{A\pi_A(X) e_D(X,0) - \frac{AD\pi_A(X)e_D(X,0)}{e_D(X,1)}\right\}\widehat{\tilde{m}}_1(r_{\alpha,1},X, S) \\
\end{aligned}\label{eq:I_1:part3}\\
     & -\left\{
     \frac{1}{|\mathcal{I}_2|}\sum_{i\in \mathcal{I}_2} D_i(1-A_i) - P(D=1, A=0)
     \right\}(1-\alpha)\label{eq:I_1:part4}\\
     & = R_1^{(S2)}(r_{\alpha,1}) + R_2^{(S2)}(r_{\alpha,1}) + R_3^{(S2)}(r_{\alpha,1}) + R_4^{(S2)}. \nonumber
\end{align}
\paragraph{Bound on $\sup_{r_{\alpha,1}}|R_1^{(S2)}(r_{\alpha,1})|$} We define a class of functions $\mathcal{F}_1$ by
$$
\mathcal{F}_1 =
\left\{f_r: 
f_r=\frac{AD\widehat{\pi}_A(X)\widehat{e}_D(X,0)\mathbf{1}(R_1 < r_{\alpha, 1})}{\widehat{e}_D(X,1)},\forall r
\right\}.
$$
One can observe that $\forall f_r\in \mathcal{F}_1$, we have $|f_r|\leq AD\pi_{0} \overline{e}_{0} \underline{e}_{0}$. Therefore, $AD\pi_{0} \overline{e}_{0} \underline{e}_{0}$ is an envelope function of $\mathcal{F}_1$. Let $\|z\|_\mathcal{F}$ denote the supermum norm of $z$ over a class of function $\mathcal{F}$, defined by $\|z\|_{\mathcal{F}} = \sup_{f\in \mathcal{F}}|z(f)|$. For $\mathcal{O}_i= (X_i, A_i, S_i, Y_i, D_i)$ and any function $f$, we define 
$$
\mathbb{G}_N = 
\frac{1}{\sqrt{N}}\sum_{i=1}^N
[f(\mathcal{O}_i) - \mathbb{P}\{f(\mathcal{O})\}].
$$
Applying Lemma \ref{lem:bound_Gn} with $s(a,d,w)=AD\widehat{\pi}_A(X)\widehat{e}_D(X,0)/\widehat{e}_D(X,1)$ and $h(w,y)=R_1$ produces $E\|\mathbb{G}_N\|_{\mathcal{F}_1}\leq C_1 \pi_{0}\overline{e}_{0} \underline{e}_{0}^{-1}$ with $C_1$ being a universal constant. By McDiarmid's inequality, we can show that
\begin{align*}
&P(\|\mathbb{G}_{N}\|_{\mathcal{F}_1}-E\|\mathbb{G}_N\|_{\mathcal{F}_1}\geq t)\leq 
\exp \left(-\frac{2t^2}{\sum_{i=1}^N c_i^2}\right)
\\
&\leq \exp \left(-\frac{2t^2}{\sum_{i=1}^N 4\pi_{0}^2\overline{e}_{0}^2\underline{e}_{0}^{-2}/N}\right) =
\exp \left(-\frac{t^2}{2\pi_{0}^2\overline{e}_{0}^2\underline{e}_{0}^{-2}}\right),  
\end{align*}
where 
\begin{align*}
    c_i &\leq \sup_{\mathcal{O}_i, \mathcal{O}_i'}\sup_{r}
\sqrt{N}\left|
\frac{1}{N}\frac{A_i D_i \widehat{\pi}_A(X_i)\widehat{e}_D(X_i,0)}{\widehat{e}_D(X_i, 1)}\mathbf{1}(R_{1,i}<r)\right. \\
&\left.- 
\frac{1}{N}\frac{A_{i'} D_{i'} \widehat{\pi}_A(X_{i'})\widehat{e}_D(X_{i'},0)}{\widehat{e}_D(X_{i'}, 1)}\mathbf{1}(R_{1,{i'}}<r)
\right| \leq \frac{2\pi_{0}\overline{e}_{0}\underline{e}_{0}^{-1}}{\sqrt{N}}.
\end{align*}
Substituting the expectation bound to the inequality gives us
\begin{equation}
P\left(\|\mathbb{G}_{N}\|_{\mathcal{F}_1}\geq C_2 \pi_{0}\overline{e}_{0}\underline{e}_{0}^{-1}\sqrt{1+\log\left(
\frac{1}{\delta}
\right)}\right)
\leq \delta, \label{eq:bound_part1}
\end{equation}
for some constant $C_2$.

\paragraph{Bound on $\sup_{r_{\alpha,1}}|R_2^{(S2)}(r_{\alpha,1})|$} Similar to the bound on $\sup_{r_{\alpha,1}}|R_1^{(S2)}(r_{\alpha,1})|$, a class of function $\mathcal{F}_2$ is defined by $\mathcal{F}_2 = \left\{
f_r: f_r = \left\{D(1-A)- 
    A\widehat{\pi}_A(X) \widehat{e}_D(X,0)\right\}\widehat{m}_1(r_{\alpha,1},X),\forall r
\right\}$, where 
\begin{align*}
    f_r &= \left\{D(1-A)- 
    A\widehat{\pi}_A(X) \widehat{e}_D(X,0)\right\}\widehat{m}_1(r_{\alpha,1},X)\\ &=  \left\{D(1-A)- 
    A\widehat{\pi}_A(X) \widehat{e}_D(X,0)\right\}
    \int_0^{m_{0}}
    \mathbf{1}\{\widehat{m}_1(r_{\alpha,1},X)\geq u\}du\\
    &= \int_0^{m_{0}} \left\{D(1-A)- 
    A\widehat{\pi}_A(X) \widehat{e}_D(X,0)\right\}
    \mathbf{1}\{r_{\alpha,1}\geq h(X, u)\}du,
\end{align*}
where the first equality holds by $\widehat{m}_1(r_{\alpha,1},X) =  \int_0^{m_{0}}
\mathbf{1}\{\widehat{m}_1(r_{\alpha,1},X)\geq u\}du$, and the second equality holds as $\widehat{m}_1(r_{\alpha,1},X)$ is monotonously increase in $r_{\alpha,1}$. By Lemma \ref{lem:bound_Gn} and McDiarmid’s inequality, we have
\begin{align*}
&E\|\mathbb{G}_N\|_{\mathcal{F}_2}\lesssim\int_0^{m_{0}}\max\{1, \pi_{0}\overline{e}_{0}\}du \lesssim\pi_{0}\overline{e}_{0}m_{0},\\
& P(\|\mathbb{G}_{N}\|_{\mathcal{F}_2}-E\|\mathbb{G}_N\|_{\mathcal{F}_2}\geq t) \leq 
\exp \left(
-\frac{t^2}{2\pi_{0}^2 \overline{e}_{0}^2m_{0}^2}
\right).
\end{align*}
Substituting the expectation bound back gives us
\begin{equation}
P\left(\|\mathbb{G}_{N}\|_{\mathcal{F}_2}\geq C_3 \pi_{0}\overline{e}_{0}m_{0}\sqrt{1+\log\left(
\frac{1}{\delta}
\right)}\right)
\leq \delta, \label{eq:bound_part2}
\end{equation}
for some constant $C_3$.

\paragraph{Bound on $\sup_{r_{\alpha,1}}|R_3^{(S2)}(r_{\alpha,1})|$}
Similarly, we first define the class of function $\mathcal{F}_3$ as
$$
\mathcal{F}_3 = \left\{f_r: \left\{A\pi_A(X) e_D(X,0) - \frac{AD\pi_A(X)e_D(X,0)}{e_D(X,1)}\right\}\widehat{\tilde{m}}_1(r_{\alpha,1},X, S)\right\}.
$$
The supermum norm of $\mathbb{G}_N$ over $\mathcal{F}_3$ and the bound of $\|\mathbb{G}_N\|_{\mathcal{F}_3}$ can be obtained by Lemma \ref{lem:bound_Gn} and McDiarmid’s inequality, respectively:
\begin{align*}
& E\|\mathbb{G}_N\|_{\mathcal{F}_3}\lesssim \int_0^{\tilde{m}_{0}} 
    \pi_{0}\overline{e}_{0}(1- \underline{e}_{0}^{-1}) du 
    \lesssim  \pi_{0}\overline{e}_{0}\underline{e}_{0}^{-1} \tilde{m}_{0},\\
    & P(\|\mathbb{G}_{N}\|_{\mathcal{F}_3}-E\|\mathbb{G}_N\|_{\mathcal{F}_3}\geq t) \leq 
\exp \left(
-\frac{t^2}{2\pi_{0}^2\overline{e}_{0}^2\underline{e}_{0}^{-2} \tilde{m}_{0}^2}
\right),
\end{align*}
which yields 
\begin{equation}
P\left(\|\mathbb{G}_{N}\|_{\mathcal{F}_3}\geq C_4 \pi_{0}\overline{e}_{0}\underline{e}_{0}^{-1} \tilde{m}_{0} \sqrt{1+\log\left(
\frac{1}{\delta}
\right)}\right)
\leq \delta,\label{eq:bound_part3}
\end{equation}
for some constant $C_4$.

\paragraph{Bound on $\sup_{r_{\alpha,1}}|R_4^{(S2)}|$}
The random variables $D_i(1-A_i)$ is i.i.d., and Hoeffding inequality gives us
$$
P(\left\{
     \frac{1}{N}\sum_{i=1}^N D_i(1-A_i) - P(D=1, A=0)
     \right\}\geq t)\leq \exp\left(-\frac{2t^2}{N}\right)
$$
which leads to
\begin{equation}
    P\left(R_4^{(S2)} \geq (1-\alpha)\sqrt{\frac{1}{2N}\log\left(\frac{1}{\delta}\right)}\right)\leq \delta. \label{eq:bound_part4}
\end{equation}
Combining (\ref{eq:bound_part1}), (\ref{eq:bound_part2}), (\ref{eq:bound_part3}), and (\ref{eq:bound_part4}) together with the help of union bound, we can show that with probability larger than $1-\delta$,
$$
\sup_{r}|R_1^{(S2)}(r_{\alpha,1}) + R_2^{(S2)}(r_{\alpha,1}) + R_3^{(S2)}(r_{\alpha,1}) + R_4^{(S2)}| \lesssim 
\pi_{0}\overline{e}_{0}(\underline{e}_{0}^{-1}+ m_{0} +\underline{e}_{0}^{-1}\tilde{m}_{0})\sqrt{\frac{\log(1/\delta) + 1}{|\mathcal{I}_2|}},
$$
which completes the proof for the bound of $I_1$. Analogously, we can show that with probability larger than $1-\delta$, 
$$
\sup_{r}|R_1^{(S1)}(r_{\alpha,1}) + R_2^{(S1)}(r_{\alpha,1}) + R_3^{(S1)}| \lesssim 
\pi_{0}\overline{e}_{0}\underline{e}_{0}^{-1}(1+ m_{0})\sqrt{\frac{\log(1/\delta) + 1}{|\mathcal{I}_2|}}
$$
holds for $\mathbb{P}\{\widehat{\psi}^{(S1)}_1(r_{\alpha,1},W)\} - \mathbb{P}_{\mathcal{I}_2}\{\widehat{\psi}^{(S1)}_1(r_{\alpha,1},W)\}$.

\subsubsection{Additional lemmas and theorems}
\begin{lemma}[Lemma 8, \cite{yang2024doubly}]
\label{lem:bound_Gn}
    There exists a universal constant $C$ such that for any functions $s(a,d,w) \in [-\kappa_0, \kappa_0]$ and $h(w,y)$,
\[
E \left[ \sup_{r} \left| \mathbb{G}_N [s(a,d,w) \mathbf{1}\{h(w,y) \leq r\}] \right| \right] \leq C \kappa_0.
\]
\end{lemma}

\begin{theorem}\label{thm:EIF0}
Under Assumptions \ref{assump:overlap}-\ref{assump:unconfoundedness}, the EIF $\psi_0^{(j)}$ for $r_{\alpha,0}$ under Setting $j$, for $j=S1, S2, S3$, is given, up to a proportionality constant, by
\begin{align*}
\psi_0^{(S1)}(r_{\alpha,0}, W;m,e_D,\pi_A)&=\psi_0^{(S3)}(r_{\alpha,0},W;m,e_D,\pi_A) =
    DA\{m_0(r_{\alpha,0}, X) - (1 - \alpha)\} \\
     &+ \frac{(1-A)De_D(X,1)}{\pi_A(X)e_D(X,0)}
    \{\mathbf{1}(R_0<{r}_{\alpha,0}) - m_0(r_{\alpha,0}, X)\},
\end{align*}
and
\begin{align*}
\psi_0^{(S2)}(r_{\alpha,0}, W;m,\Tilde{m},e_D,\pi_A) &=  
    DA\{m_0(r_{\alpha,0}, X) - (1 - \alpha)\} \\
    & + \frac{(1-A)e_D(X,1)}{\pi_A(X)}\{\Tilde{m}_0(r_{\alpha,0}, X,S) - m_0(r_{\alpha,0}, X)\}  \\
    & + \frac{(1-A)De_D(X,1)}{\pi_A(X)e_D(X,0)}
    \{\mathbf{1}(R_1<{r}_{\alpha,0}) - \Tilde{m}_0(r_{\alpha,0}, X,S)\},
\end{align*}
where $\Tilde{m}_0(r_{\alpha,0}, X,S)=P(R_0<{r}_{\alpha,0}\mid X,S,A=0,D=1)$ , $m_0(r_{\alpha,0}, X) = P(R_0<r_{\alpha,0}\mid X, A=0, D=1)$.
\end{theorem}

\begin{corollary}\label{coro:v0_reduction}
    Under Assumptions \ref{assump:overlap}-\ref{assump:unconfoundedness}, the efficiency gain from observing surrogates in both datasets (Setting 2) compared to only in the source data (Setting 1) in estimating $r_{\alpha,0}$ is given by
    \begin{align*}
        & V_0^{(S2)} - V_0^{(S1)} = 
        E \left[
        \frac{1 - e_D(X,0)}{e_D(X,0)}
        \frac{e_D^2(X,1)\{e_A(X)\}^2}{1-e_A(X)}
        \var \{\Tilde{m}_0(r_{\alpha,0},X,S)\mid X, A=0\}
        \right],
    \end{align*}
    where 
    \begin{align*}
        V_0^{(S1)} = 
    \var\{\psi_0^{(S1)}(r_{\alpha,0}, W;m,e_D,\pi_A)\}, \quad V_0^{(S2)} = 
    \var\{\psi_0^{(S2)}(r_{\alpha,0}, W;m,\Tilde{m},e_D,\pi_A)\}.
    \end{align*}
\end{corollary}

\newpage

\section{Additional Simulation Studies}\label{more:sims}

In this section, we assess the group-conditional performance of the prediction intervals produced by the SCIENCE framework for categorical outcomes. The data generation procedures for $X$, $A$, and $S$ remain the same as before, while the potential primary outcomes $Y(a)$ are generated with five levels under the multinomial distribution as follows:
$$
\frac{P(Y(a) = k \mid X, S)}{P(Y(a) = 1 \mid X, S)} = \exp\left(-\alpha_{Y(a),k} - \sum_{j=1}^2 \frac{X_j}{2} - \sum_{j=1}^2 \frac{S_j(a)}{2}\right)
$$
for $k = 2, \dots, 5$, where $k = 1$ serves as the reference level. The parameters $\alpha_{Y(a),k}$ are adaptively chosen to ensure the marginal probabilities are approximately $P(Y(1) = k) = (0.1, 0.2, 0.4, 0.15, 0.15)$ and $P(Y(0) = k) = (0.3, 0.3, 0.2, 0.15, 0.05)$ for $k = 1, \dots, 5$.

The non-conformity scores for categorical outcomes are derived using the nested prediction sets approach from \cite{kuchibhotla2023nested}. Let $F_a(W; t) = \{y : P_{Y(a)}(y \mid W) \leq t\}$ denote a nested sequence of sets constructed based on the first data fold, $\mathcal{I}_1$. The associated non-conformity scores $R(W, Y(a))$ are then defined as $R(W, Y(a)) = \inf \{t : y \in F_a(W; t)\}$, where the conditional probabilities are estimated using multinomial logistic regression from the \texttt{nnet} package, with or without the inclusion of surrogates. The semi-parametric efficient estimator for the $(1 - \alpha)$-quantile $r_\alpha$ of $R(W, Y(a))$ is then computed using the second data fold, $\mathcal{I}_2$.

\begin{figure}[!ht]
    \centering
    \includegraphics[width=.85\linewidth]{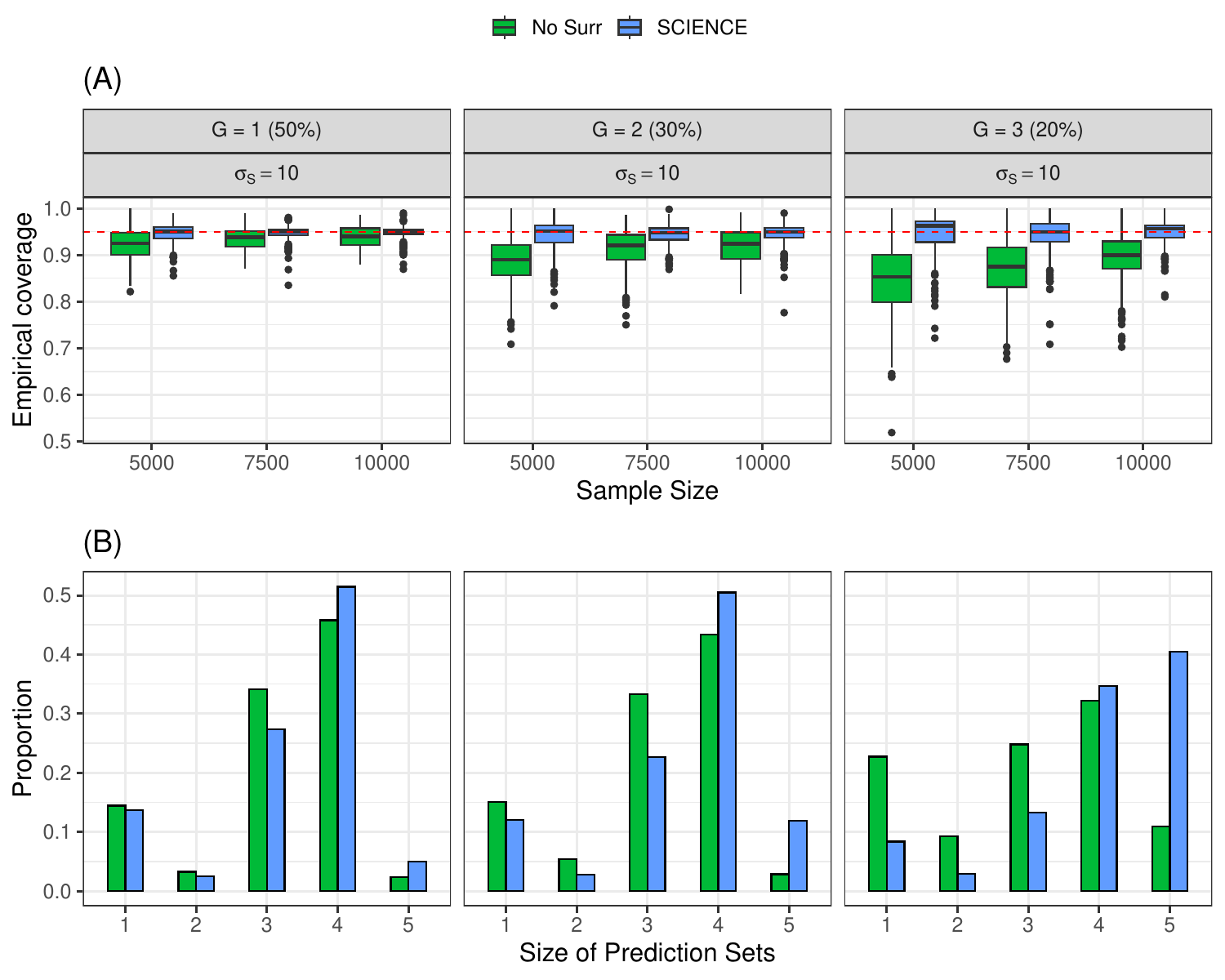}
    \caption{(A) Empirical coverage of the 95\% prediction intervals, and (B) the proportions of the size of prediction sets for the ITE of the combined data when $\sigma_S = 10$ and the outcomes are categorical, conditional on the group variable $G$, across 500 replicates.}
    \label{fig:rstDwR_combined}
\end{figure}

Figure \ref{fig:rstDwR_combined}(A) presents the empirical coverage of the prediction sets at $\alpha=0.05$ for the observed primary outcomes of both source and target datasets, conditional on the group variable $G$. The SCIENCE framework that adjusts for the effect of surrogates on the conditional probabilities yields valid empirical coverage for the prediction sets. In contrast, the efficient estimator without surrogates fails to achieve valid empirical coverage, as the conditional probability model marginalized over the surrogates no longer admits the multinomial logistic specification. This misspecification leads to incorrectly constructed non-conformity scores and therefore poor empirical coverage. Figure \ref{fig:rstDwR_combined}(B) provides the proportions of the size of prediction sets, varying from $1$ to $5$, conditional on the group variable $G$. While the prediction sets produced by the efficient estimator without surrogates tend to be smaller than the SCIENCE framework, these smaller sets can be misleading, as they often fail to contain the true primary outcomes, indicated by their poor coverage.

\newpage

\section{Additional Real Data Analysis}\label{more:real-data}

\subsection{Coverage and Width for Target and Source Baseline SARS-CoV-2 Seropositive Participants}

\begin{table}[htbp!]
\centering
\caption{Baseline SARS-CoV-2 Seropositive participants: Average coverage and prediction interval width on the source (top panel) and target (bottom panel) population over $100$ repetitions at different miscoverage levels $\alpha$, comparing WCQR \citep{lei2021conformal}, our method without surrogates (No Surr), SCIENCE using Day 29 binding spike antibody level as the surrogate (Binding), SCIENCE using Day 29 nAb-50 as the surrogate (nAb-50), and SCIENCE using both surrogates (Both).}
\vspace{0.5em}
\resizebox{\textwidth}{!}{%
\begin{tabular}{c|ccccc|ccccc}
\toprule
\textbf{$\alpha$} & \multicolumn{5}{c|}{\textbf{Coverage $\times 100\%$}} & \multicolumn{5}{c}{\textbf{Prediction Interval Width}} \\
\cmidrule(lr){2-6} \cmidrule(lr){7-11}
 & \textbf{WCQR} & \textbf{No Surr} & \textbf{SCIENCE} & \textbf{SCIENCE} & \textbf{SCIENCE} & \textbf{WCQR} & \textbf{No Surr} & \textbf{SCIENCE} & \textbf{SCIENCE} & \textbf{SCIENCE} \\
 &  &  & \textbf{(Binding)} & \textbf{(nAb-50)} & \textbf{(Both)} &  &  & \textbf{(Binding)} & \textbf{(nAb-50)} & \textbf{(Both)} \\
\midrule
0.05 & 95.2 & 95.3 & 94.9  & 95.5 & 95.8 &  1.98 & 1.91 & 1.83 & 1.82 & 1.72 \\
0.10 & 90.1 & 90.0 & 89.7 & 90.3 & 90.6  & 1.63 & 1.51 & 1.44 & 1.42 & 1.40 \\
0.20 & 81.1 & 79.8 & 79.0 & 80.6 & 82.5 & 1.26 & 1.21 & 1.16 & 1.15 & 1.13 \\
0.30 & 72.4 & 70.0 & 69.9 & 70.2 & 72.7 & 1.04 & 1.01 & 1.00 & 1.00 & 0.98 \\
0.40 & 56.7 & 60.0 & 60.0 & 59.2 & 60.4 & 0.88 & 0.88 & 0.87 & 0.85 & 0.83 \\
\bottomrule
\end{tabular}
}
\resizebox{\textwidth}{!}{%
\begin{tabular}{c|ccccc|ccccc}
\\
\toprule
\textbf{$\alpha$} & \multicolumn{5}{c|}{\textbf{Coverage $\times 100\%$}} & \multicolumn{5}{c}{\textbf{Prediction Interval Width}} \\
\cmidrule(lr){2-6} \cmidrule(lr){7-11}
 & \textbf{WCQR} & \textbf{No Surr} & \textbf{SCIENCE} & \textbf{SCIENCE} & \textbf{SCIENCE} & \textbf{WCQR} & \textbf{No Surr} & \textbf{SCIENCE} & \textbf{SCIENCE} & \textbf{SCIENCE} \\
 &  &  & \textbf{(Binding)} & \textbf{(nAb-50)} & \textbf{(Both)} &  &  & \textbf{(Binding)} & \textbf{(nAb-50)} & \textbf{(Both)} \\
\midrule
0.05 & 94.7 & 95.8 & 94.7 & 94.8 & 94.7 & 2.14 & 2.02 & 1.92 & 1.88 & 1.77 \\
0.10 & 90.9 & 90.6 & 90.2 & 92.7 & 92.7 & 1.73 & 1.59 & 1.51 & 1.50 & 1.47 \\
0.20 & 82.5 & 81.7 & 80.9 & 82.5 & 83.0  & 1.30 & 1.26 & 1.20 & 1.20 & 1.16 \\ 
0.30 & 72.7 & 70.6 & 71.0 & 73.2 & 73.4 & 1.07 & 1.04 & 1.03 & 1.02 & 0.99 \\
0.40 & 55.9 & 58.2 & 61.1 & 65.3 & 65.8 &  0.90 & 0.90 & 0.88 & 0.86 & 0.84 \\
\bottomrule
\end{tabular}
}
\label{maintab:cp_width_source_target}
\end{table}

\subsection{Baseline Seronegative Participants}

We repeated the main data analysis for a subcohort of individuals who were SARS-CoV-2 seronegative at baseline. A total of 1,188 participants were included in this sample. Among these participants, 140 were placebo recipients ($A=0$), and 1,048 received the two-dose mRNA-1273 vaccine $(A=1)$. We designated the target population as underrepresented minorities $n_{D0} = 636$ $(53.5\%)$, defined as Blacks or African Americans, Hispanics or Latinos, American Indians or Alaska Natives, Native Hawaiians, and other Pacific Islanders. The source population of non-minorities $n_{D1} = 532$ $(46.5\%)$ included all other races with observed race (Asian, Multiracial, White, Other) and observed ethnicity not being Hispanic or Latino.

We used a 75-25 split for training and testing, and to account for the variability from data splitting, we repeated the above procedure 100 times and summarized the average coverage and widths of prediction intervals of the observed outcomes for different methods. Our results show that SCIENCE produces prediction intervals with nominal coverage across the combined target and source populations for various miscoverage levels $\alpha \in \{0.05, 0.1, 0.2, 0.3, 0.4\}$ (see Table~\ref{tab:coverage_and_width_overall}). In terms of average prediction interval width for the ITE, summarized in the same table, WCQR yields the widest intervals. SCIENCE generates shorter average intervals than the efficient method without surrogates. Using the Day 29 nAb-ID50 titer as a surrogate results in narrower intervals compared to the Day 29 binding antibody titer alone, with the shortest intervals achieved when both Day 29 markers are used. As expected, prediction intervals for the target population are slightly wider due to the nested conformal inference procedure, and intervals increase in width as $\alpha$ decreases. Detailed results for the source and target populations are separately provided in Table~\ref{tab:cp_width_source_target}.

\begin{table}[htbp]
\label{tab:coverage_and_width_overall}
\centering
\caption{Seronegative participants: Average coverage and prediction interval width across the combined population of target and source test points over $100$ repetitions at different miscoverage levels $\alpha$, comparing WCQR \citep{lei2021conformal}, efficient method without surrogates (No Surr), SCIENCE using Day 29 binding spike antibody level as the surrogate (Binding), SCIENCE using Day 29 nAb-ID50 titer as the surrogate (nAb-50), and SCIENCE using both surrogates (Both).}
\vspace{0.5em}
\resizebox{\textwidth}{!}{
\begin{tabular}{c|ccccc|ccccc}
\toprule
\textbf{$\alpha$} & \multicolumn{5}{c|}{\textbf{Coverage $\times 100\%$}} & \multicolumn{5}{c}{\textbf{Prediction Interval Width}} \\
\cmidrule(lr){2-6} \cmidrule(lr){7-11}
 & \textbf{WCQR} & \textbf{No Surr} & \textbf{SCIENCE} & \textbf{SCIENCE} & \textbf{SCIENCE} & \textbf{WCQR} & \textbf{No Surr} & \textbf{SCIENCE} & \textbf{SCIENCE} & \textbf{SCIENCE} \\
  &  & & \textbf{(Binding)} & \textbf{(nAb-50)} & \textbf{(Both)} & &  & \textbf{(Binding)} & \textbf{(nAb-50)} & \textbf{(Both)} \\
\midrule
0.05 & 94.9 & 94.8 & 95.5 & 95.2 & 95.3 & 1.54 & 1.49 & 1.43 & 1.34 & 1.33 \\
0.10 & 91.7 & 90.8 & 91.3 & 91.8 & 91.7 & 1.23 & 1.19 & 1.14 & 1.12 & 1.11 \\
0.20 & 80.8 & 80.5 & 81.3 & 82.1 & 82.3 & 0.89 & 0.82 & 0.80 & 0.81 & 0.81 \\
0.30 & 72.0 & 70.8 & 71.8 & 73.0 & 73.0 & 0.69 & 0.65 & 0.64 & 0.65 & 0.65 \\
0.40 & 63.7 & 62.7 & 63.7 & 64.4 & 64.7 & 0.60 & 0.56 & 0.56 & 0.54 & 0.54 \\
\bottomrule
\end{tabular}
 }
\label{tab:cp_width_SM}
\end{table}

\begin{figure}[!htbp]
    \centering
    \includegraphics[width=\linewidth]{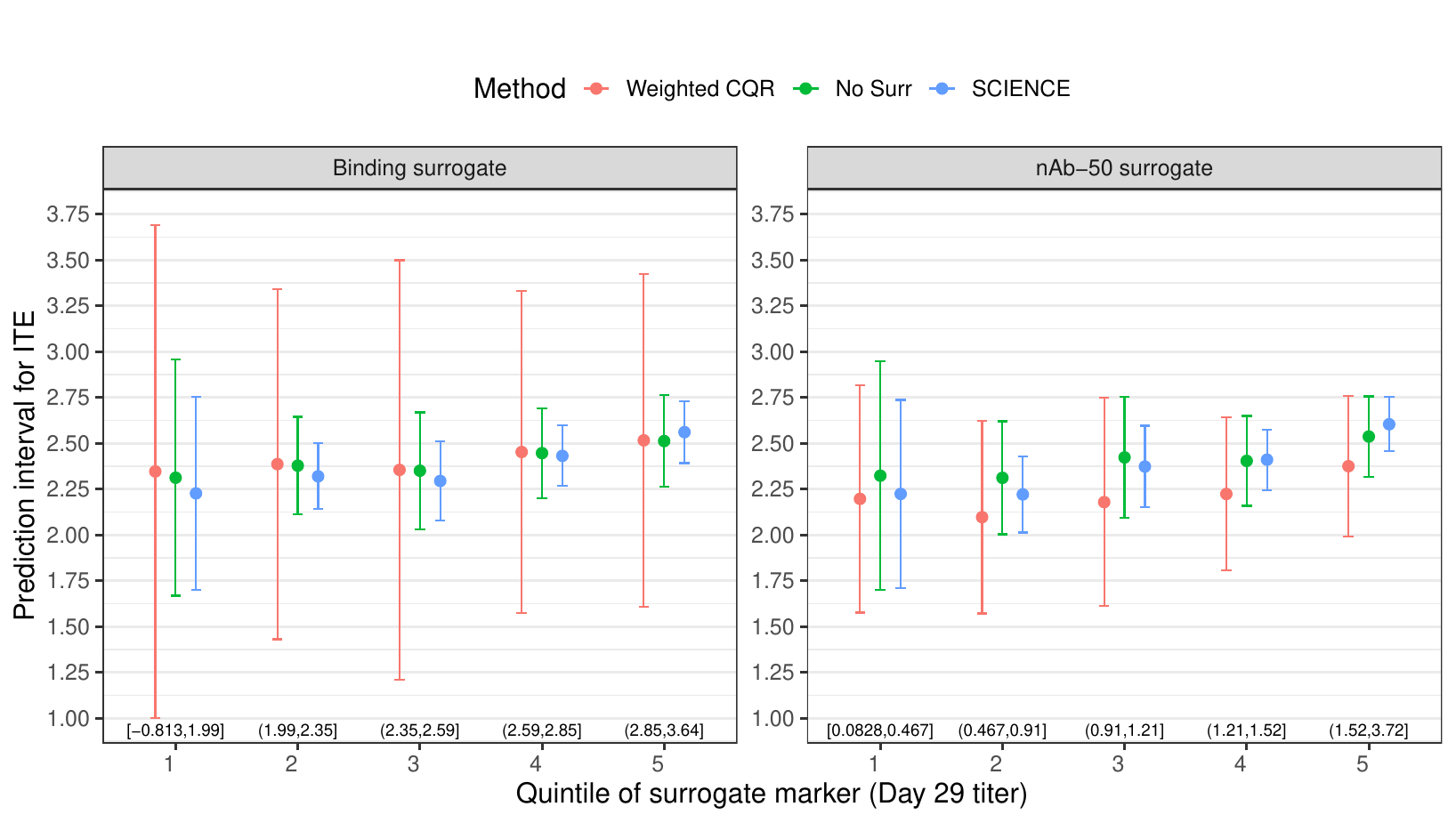}
    \caption{$90\%$ prediction intervals for the ITE on the Day 57 neutralizing antibody titer primary outcome across quintiles of Day 29 surrogate marker titers}
    \label{fig:pred-int}
\end{figure}

To visualize the efficiency gain from incorporating surrogate markers, we plotted the 90\% prediction intervals for the ITE on log$_{10}$ Day 57 neutralizing antibody titer, defined as the difference in log$_{10}$ titers for individuals who received the two-dose mRNA-1273 vaccine versus those who received a placebo. The prediction intervals were calculated across five levels of surrogate marker values: very low, low, medium, high, and very high. The quintiles were computed pooling over vaccine and placebo arms. Within each quintile, the lower and upper bounds of the prediction intervals were averaged across individuals. We focused on Setting 2, incorporating the two surrogates, while also considering Setting 1, represented by the Weighted CQR and No Surr estimators. As shown in Figure~\ref{fig:pred-int}, the lower bounds of the intervals around the ITE were consistently positive across all groups, indicating a favorable vaccine effect relative to placebo. Furthermore, the prediction intervals shifted to higher values for individuals with higher surrogate marker levels, suggesting that higher Day 29 binding antibody titer and nAb-ID50 titer predicted a larger vaccine effect on the Day 57 titer. Notably, SCIENCE yielded the narrowest prediction intervals, demonstrating the efficiency gain achieved by incorporating surrogate information.

\begin{table}[htbp]
\centering
\caption{Seronegative participants: Average coverage and prediction interval width on the source (top panel) and target (bottom panel) population over $100$ repetitions at different miscoverage levels $\alpha$, comparing WCQR \citep{lei2021conformal}, our method without surrogates (No Surr), SCIENCE using Day 29 binding spike antibody level as the surrogate (Binding), SCIENCE using Day 29 nAb-50 as the surrogate (nAb-50), and SCIENCE using both surrogates (Both).}
\vspace{0.5em}
\resizebox{\textwidth}{!}{%
\begin{tabular}{c|ccccc|ccccc}
\toprule
\textbf{$\alpha$} & \multicolumn{5}{c|}{\textbf{Coverage $\times 100\%$}} & \multicolumn{5}{c}{\textbf{Prediction Interval Width}} \\
\cmidrule(lr){2-6} \cmidrule(lr){7-11}
 & \textbf{WCQR} & \textbf{No Surr} & \textbf{SCIENCE} & \textbf{SCIENCE} & \textbf{SCIENCE} & \textbf{WCQR} & \textbf{No Surr} & \textbf{SCIENCE} & \textbf{SCIENCE} & \textbf{SCIENCE} \\
 &  &  & \textbf{(Binding)} & \textbf{(nAb-50)} & \textbf{(Both)} &  &  & \textbf{(Binding)} & \textbf{(nAb-50)} & \textbf{(Both)} \\
\midrule
0.05 & 95.0 & 95.7 & 96.0 & 95.6 & 95.8 & 1.52 & 1.46 & 1.41 & 1.32 & 1.32 \\
0.10 & 92.7 & 91.2 & 91.0 & 91.2 & 90.9 & 1.22 & 1.17 & 1.13 & 1.11 & 1.10 \\
0.20 & 82.0 & 82.5 & 81.8 & 82.3 & 81.7 & 0.87 & 0.81 & 0.79 & 0.81 & 0.81 \\
0.30 & 75.6 & 73.8 & 72.9 & 73.5 & 72.8 & 0.68 & 0.64 & 0.63 & 0.65 & 0.65 \\
0.40 & 66.1 & 65.1 & 64.2 & 64.8 & 63.8 & 0.59 & 0.55 & 0.55 & 0.54 & 0.53 \\
\bottomrule
\end{tabular}
}
\resizebox{\textwidth}{!}{%
\begin{tabular}{c|ccccc|ccccc}
\\
\toprule
\textbf{$\alpha$} & \multicolumn{5}{c|}{\textbf{Coverage $\times 100\%$}} & \multicolumn{5}{c}{\textbf{Prediction Interval Width}} \\
\cmidrule(lr){2-6} \cmidrule(lr){7-11}
 & \textbf{WCQR} & \textbf{No Surr} & \textbf{SCIENCE} & \textbf{SCIENCE} & \textbf{SCIENCE} & \textbf{WCQR} & \textbf{No Surr} & \textbf{SCIENCE} & \textbf{SCIENCE} & \textbf{SCIENCE} \\
 &  &  & \textbf{(Binding)} & \textbf{(nAb-50)} & \textbf{(Both)} &  &  & \textbf{(Binding)} & \textbf{(nAb-50)} & \textbf{(Both)} \\
\midrule
0.05 & 93.5 & 93.9 & 95.4 & 94.8 & 94.7 & 1.57 & 1.51 & 1.45 & 1.36 & 1.34 \\
0.10 & 91.5 & 90.3 & 91.6 & 92.7 & 92.7 & 1.25 & 1.20 & 1.15 & 1.12 & 1.11 \\
0.20 & 78.1 & 78.4 & 80.9 & 82.5 & 83.0 & 0.90 & 0.84 & 0.81 & 0.81 & 0.82 \\
0.30 & 72.7 & 67.7 & 70.7 & 73.2 & 73.4 & 0.70 & 0.66 & 0.65 & 0.66 & 0.66 \\
0.40 & 65.7 & 60.2 & 63.3 & 65.3 & 65.8 & 0.61 & 0.57 & 0.57 & 0.55 & 0.55 \\
\bottomrule
\end{tabular}
}
\label{tab:cp_width_source_target}
\end{table}

\end{document}